\documentclass[a4paper]{article}

\usepackage{lineno,hyperref,psfrag,latexsym,amssymb}
\usepackage{graphicx,color}

\modulolinenumbers[5]
\renewcommand{\vec}[1]{\mbox{\boldmath $#1$}} 

\def\bra{\langle}
\def\ket{\rangle}
\def\p{\partial}

\def\beq{\begin{equation}}
\def\eeq{\end{equation}}
\def\la{\label}

\def\degree{$^{\rm o}$}

\newcommand{\mylab}[3]{\raisebox{#2}[0mm][0mm]{%
\makebox[0mm][l]{\hspace*{#1}{#3}}}}%
\def\utau{u_\tau}
\def\retau{Re_\tau}
\def\diss{\varepsilon}
\def\tg{\overline{g}}
\def\uvec{\mbox{\boldmath $u$}}
\def\figpath{./}
\def\spacce#1{\hskip #1pt}
\def\drawline#1#2{\raise 2.5pt\vbox{\hrule width #1pt height #2pt}}
\def\solid{\drawline{24}{.5}\nobreak}

\def\bdash{\hbox{\drawline{5.8}{.5}\spacce{2}}}

\def\dashed{\bdash\bdash\bdash\nobreak}

\def\bdot{\hbox{\drawline{1}{.5}\spacce{2}}}

\def\dotted{\hbox{\leaders\bdot\hskip 24pt}\nobreak}

\def\circle{$\circ$\nobreak }

\def\trian{\raise 1.25pt\hbox{$\scriptstyle\triangle$}\nobreak}

\def\dtrian{\raise 1.25pt\hbox%
{$\scriptscriptstyle\bigtriangledown$}\nobreak}

\def\squar{\raise 1.25pt\hbox{$\scriptstyle\Box$}\nobreak}

\def\diamon{\raise 1.25pt\hbox{$\scriptstyle\diamond$}\nobreak}

\newcommand{\soliddtrian}{$\blacktriangledown$\nobreak}

\def\linedtri1{\hbox{\bdash\hspace{-1.6mm}$\bigtriangleup$\hspace{-0.8mm}\bdash}\nobreak}
\def\soliddtrian1{$\blacktriangledown$\nobreak}
\def\solidrtrian2{$\blacktriangleright$\nobreak}
\def\solidltrian3{$\blacktriangleleft$\nobreak}

\title{\huge\bf Computers and turbulence}

\author{\rm Javier Jim\'enez\\
School of Aeronautics, U. Polit\'ecnica Madrid\\ 28040 Madrid Spain}
\date{\today}

\begin{document}
\maketitle
\begin{abstract}
This paper briefly reviews the influence that the rapid evolution of computer power in the
last decades has had on turbulence research. It is argued that it can be divided into three
stages. In the earliest (`heroic') one, simulations were expensive and could at most be
considered as substitutes for experiments. Later, as computers grew faster and some
meaningful simulations could be performed overnight, it became practical to use them as
(`routine') tools to provide answers to specific theoretical questions. More recently,
some turbulence simulations have become trivial, able to run in minutes, and it is possible
to think of computers as `Monte Carlo' theory machines, which can be used to systematically
pose a wide range of `random' theoretical questions, only to later evaluate which of them
are interesting or useful. Although apparently wasteful, it is argued that this procedure
has the advantage of being reasonably independent of received wisdom, and thus more able
than human researchers to scape established paradigms. The rate of growth of computer power
ensures that the interval between consecutive stages is about fifteen years. Rather than
offering conclusions, the purpose of the paper is to stimulate discussion on whether
machine- and human-generated theories can be considered comparable concepts, and on how the
challenges and opportunities created by our new computer `colleagues' can be made to fit
into the traditional research process.
\end{abstract}



\thispagestyle{myheadings}\markright{{\it European J. Mech.: B Fluids} {\bf 79}, \rm 1--11 (2019)}
\section{Introduction}\la{sec:intro}

Although, from its title, this paper may seem to be an appeal for a new and unnecessary
journal, it is intended as a review of the role that the rapid development of direct
numerical simulations has played over the last decades in the elucidation of turbulence
physics, and as a meditation on its possible evolution in the immediate future. It does not
address the parallel development of turbulence models and large-eddy simulations, which are
the mainstay of industry but which owe less to computational power. As the story unfolds, it
should be clear to the reader that many of the themes that we meet in tracing turbulence
research can be generalised to the role of computers in other branches of science and
technology, some of which have led the way, while others have lagged behind
\cite{bren96,norm:96,moin98,Wu:Teg:18}. We may use the former as guides to likely future
developments in our field, and the latter as warnings of the pitfalls to be avoided. In
addition, and even if a definite answer is well beyond the scope of a paper like the present
one, we will briefly discuss what could be the consequences of continuing in the present
direction, both for the field and for its practitioners, and leave to the readers the
decision on what they think is the right course to follow.

It has become customary to start historical reviews of turbulence with a reference to
Leonardo da Vinci. Unfortunately, although he wrote extensively on fluid mechanics and on
what later came to be regarded as turbulence, I have been unable to trace any reference to
computation or automata applicable to the present review. However, to maintain tradition, I
offer a quote which is relevant to the general themes discussed in the paper.

It refers to the relation between theory, experience and data, and can be applied to the
hopes of some people, including the author, of what could be expected from computers. It can
also be taken as a working definition of a successful theory: ``There is nothing in nature
without a cause; understand the cause and you will have no need for the experiment''
\cite[Cod. Atlantico, 147va]{MacC:1939}. This is clearly a dubious statement, and Leonardo
hedges it elsewhere by declaring that: ``It is my intention first to cite experience, then
to demonstrate through reasoning why experience operates in a given way'' \cite[Ms E Paris,
55r]{Zamm:80}, or that we should ``avoid the teaching of speculators whose judgement is not
confirmed by experience'' \cite[Ms B Paris, 4v]{Zamm:80}. While aiming for the first of
these quotes, this review will try to follow the last two.

The paper is organised in terms of the cost of individual simulations. Section
\ref{sec:simul} deals with simulations which are very expensive, and
\S\ref{sec:structures} discusses the consequences of some simulations becoming inexpensive
enough to be considered routine. These two sections can be seen as a short historical
survey of the field up to now, but \S\ref{sec:curio} speculates on how turbulence research
might be affected when simulations become cheap enough to be deemed trivial, and offers an
example. The paper is intended to promote discussion among researchers, and offers no
conclusions, but a summary and a list of possible open questions are collected in
\S\ref{sec:discus}.

\section{Turbulence and numerical simulation}\la{sec:simul}

Humans must have been aware from very early times that the flow of water is not always
smooth, and that crossing a river involves dealing with eddies comparable to the mean drift
of the stream. The same applies to the wind, and some of the first recorded uses of the word
`turbulent' are applied to the weather. Quantitative study of turbulent flows, even if not
recognised as such, must also have been commonplace, because the design of the extensive
irrigation canals and aqueducts of antiquity had to rest on a correct estimation of the
friction coefficient. Both ancient Greek $(\tau\acute{\upsilon}\rho\beta\eta)$ and Latin
({\em turba--\ae}) have related words meaning disorder, confusion or turmoil. Several Latin
languages still maintain the word {\em turba} to describe an unruly mob. Latin has the
adjective {\it turbulentus--a}, applying to anything disordered or confused, as in {\it
turbulent times}, weather or sea. The adjective and the verb ({\it turbare}), rather than
the noun, were kept in use during the Middle Ages. In Spanish, Gonzalo de Berceo (circa
1230) uses them to refer to the weather, to visions and to character. In English, they don't
appear to have been introduced until the XVI century. The Oxford dictionary points to a
first use of `turbulent' in the Bible of Coverdale (1538), referring to the ``attacks of the
wicked'', and to a first reference to `turbulent weather' in the letters of G.~Harvey in
1573.

In none of these instances, including those referring to the atmosphere or to the ocean, is
there any implication that turbulence is a separate kind of flow. Leonardo da Vinci
\cite{MacC:1939,Zamm:80} wrote extensively around 1500 on hydraulic applications and
machines, and did detailed sketches and descriptions of the wakes behind objects and of
the eddying motions on the surface of pools and rivers. He worried about their effect on bed
erosion, and had multiple (mostly incorrect) theories about the origin and effects of
eddies, but most hydrodynamics up to the late XIX century was essentially laminar, and
dealt with the mean properties of the flow without regard to its fluctuations.

The first recognition of the existence of two different flow regimes probably came around
the middle of the XIX century from careful measurements of water friction in pipes. This was
at the time a classical problem in practical hydraulics which, together with that of the
drag of moving bodies, had been the subject of experimental and theoretical investigations
for at least 150 years. It was known that it had two components, one linear and the other
approximately quadratic in the fluid velocity, and that only the first one depends on
the viscosity of the fluid. Hagen \cite{hag39} and Poiseuille \cite{pois46} had inaugurated
modern rheology by studying the linear regime in capillaries, and by measuring the
dependence of the viscosity on temperature. In 1854 Hagen \cite{hag54} and Darcy
\cite{dar55} independently published careful measurements on larger pipes and noted that the
nonlinear component was dominant, and that it came together with disordered motion in the
fluid. Both speculated that the increased drag was due to the energy spent in creating the
fluctuating eddies.

Soon after that, Boussinesq \cite{bous77} published a long paper clearly distinguishing
between two different kinds of flow, smooth and `tumultuous'. It contained many of the
ideas that were later associated with turbulence research, such as a modified eddy
viscosity and the realisation that turbulent flows were too complicated for a deterministic
description, and had to be treated statistically.

The transition between the two states was clarified in the famous experiments of 
Reynolds \cite{rey83}, in which he introduced dye at the inlet of a long pipe and
observed the process by which the flow became disordered; this is the paper in which he
introduced what became known as the `Reynolds number', to describe the circumstances at
which the transition took place. In his second important paper on the subject \cite{rey94}
he used for the first time the decomposition of the flow into mean and fluctuating parts,
and introduced the concept of Reynolds stresses. It is from this paper that much of the
modern statistical theory of turbulence derives. A more complete account of this early
period of hydrodynamics can be found in \cite{rouse}.

The first recorded reference to the perturbed flow as {\it turbulent} is in the second
edition of Lamb \cite{lamb95}, who attributes the name to Lord Kelvin and warns that it was,
at the time, the ``chief outstanding difficulty'' of hydrodynamics.

Although some would argue that this warning still applies, the subject has advanced substantially
over the intervening century, driven intermittently by improvements in instrumentation and
in theory. The 1930's saw the extension to the study of turbulence of the general flowering
of physics, especially that of statistical mechanics, culminating in Kolmogorov's \cite{kol41}
prediction of the form of the energy spectrum. The following years were dominated by the
introduction of the hot wire anemometer, which provided unprecedented amounts of
experimental detail on the smaller scales of the flow, joined in the late 1960's by a
renewed interest in optical techniques, which partially shifted the emphasis towards larger
coherent eddies.

\begin{figure}[t]
\centering
%
\raisebox{0mm}{\includegraphics[height=.35\textwidth,clip]%
{\figpath 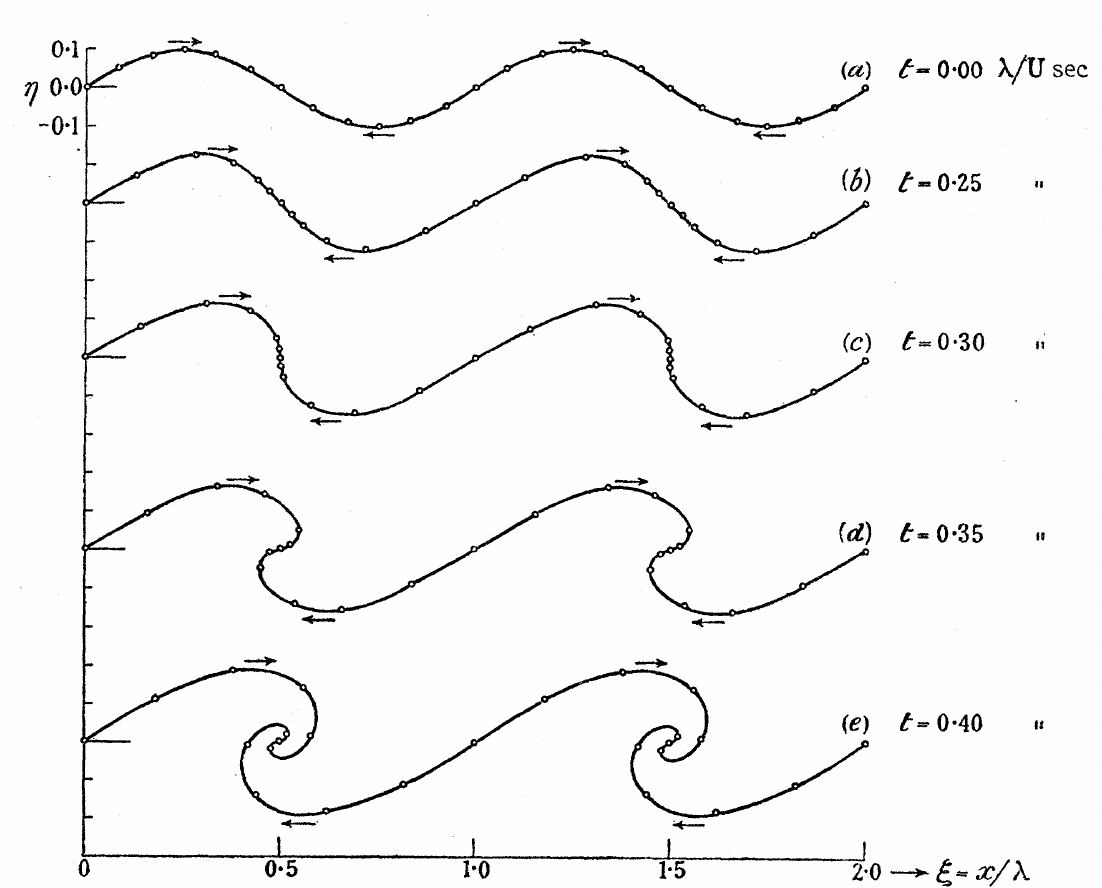}}%
\mylab{-.23\textwidth}{.37\textwidth}{(a)}%
\hspace*{5mm}%
\raisebox{0mm}{\includegraphics[height=.35\textwidth,clip]%
{\figpath 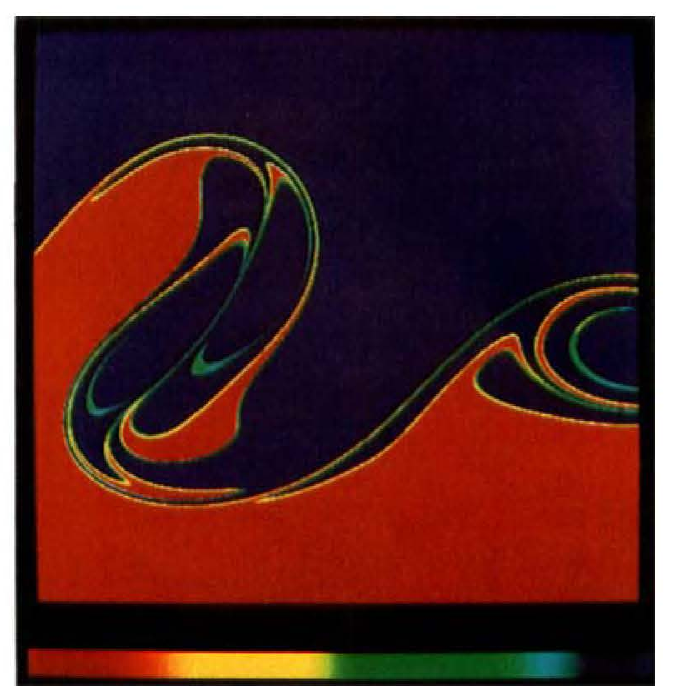}}%
\mylab{-.18\textwidth}{.37\textwidth}{(b)}%
\caption{%
(a) Numerical simulation of the roll-up of a vorticity layer. Reproduced with permission
from \cite{rosen31}.
(b) Laser-induced fluorescent picture of an experimental mixing layer in water. Reproduced
with permission from \cite{kooch:86}.
}
\label{fig:rosen}
\end{figure}

The first real contribution of computers to turbulence was due to Richardson in the early
1920's. Up to then, most simulations of turbulence had centred on the mean flow, and on
modelling averaged quantities such as drag or mixing efficiency. That was natural, because
fluctuations in engineering flows are too fast and too small to be easily
observable or relevant. But Richardson was a meteorologist, and the velocity
fluctuations in the atmosphere range from weather systems with a lifetime of days and a
length scale of the order of a continent, to wind draughts and tornadoes lasting minutes and
severely impacting human activity. The mean atmospheric circulation is of great theoretical
interest, but its velocity fluctuations are a practical problem.

Besides several seminal contributions to the observation and theory of turbulence,
Richardson proposed a way to discretise the equations of weather prediction in terms of
centred finite differences \cite{rich22}. He applied them to an actual simulation of the
weather over middle Europe and, although this first experiment did not succeed, his work is
now recognised as the foundation of modern prediction schemes \cite{hunt:97}. At that time,
a computer was a person whose role was to compute, either by hand or with mechanical aids
such as a slide rule. Richardson performed his simulation manually while serving at the
front during World War I, and he was acutely aware that many more operators would be
required to simulate the weather in less than real time. He dedicates the
last part of \cite{rich22} to the `coding' organisation of the work, with a set of forms to
be filled by individuals and passed to their neighbours, in an early implementation of
parallel multiprocessing and networking. He also estimated the number of workers
(64,000) required for a global forecast \cite{lynch:93}. He even tried to fund the scheme,
but he failed, and meteorological forecasting took a different direction until the advent of
electronic computing in the 1950's.
  
Not long afterwards, but still before the introduction of electronic computers, Rosenhead
\cite{rosen31} presented the numerical simulation of the nonlinear evolution of the
Kelvin--Helmholtz instability of a velocity discontinuity. He represented it as a series of
point vortices whose motion was traced by a Runge--Kutta scheme, and he was able to describe
the formation of the large-scale vortices that later became associated with turbulent shear
layers (see figure \ref{fig:rosen}).

Electronic computers began to appear in the 1940's and became available for scientific
research soon afterwards \cite{seid96}. John von Neumann promoted the design and
construction of the first general purpose computers, and organised several groups to exploit
them at the Princeton Institute for Advanced Studies \cite{Macrae:1999}. He foresaw that
computers would revolutionise the study of nonlinear problems, which were seldom considered
at the time for lack of appropriate tools, and identified weather prediction, aerodynamics
and turbulence as key areas of application. One of the first weather forecasts
resulting from this program is \cite{charetal50}, which follows in part the lead of
Richardson thirty years before.

The application of computers to the direct simulation of the turbulence fluctuations
proceeded quickly after that, mostly driven by the exponential increase in processor speed and
memory capacity. The same has been true of experiments, spurred in part by mutual
competition with simulations, but it is probably true that, mostly because of their superior
observational capabilities, most of the theoretical advances in fundamental turbulence
physics now come from simulations.

\begin{figure}[t]
\centering
\vspace*{2mm}%
\raisebox{0mm}{\includegraphics[height=.35\textwidth,clip]%
{\figpath 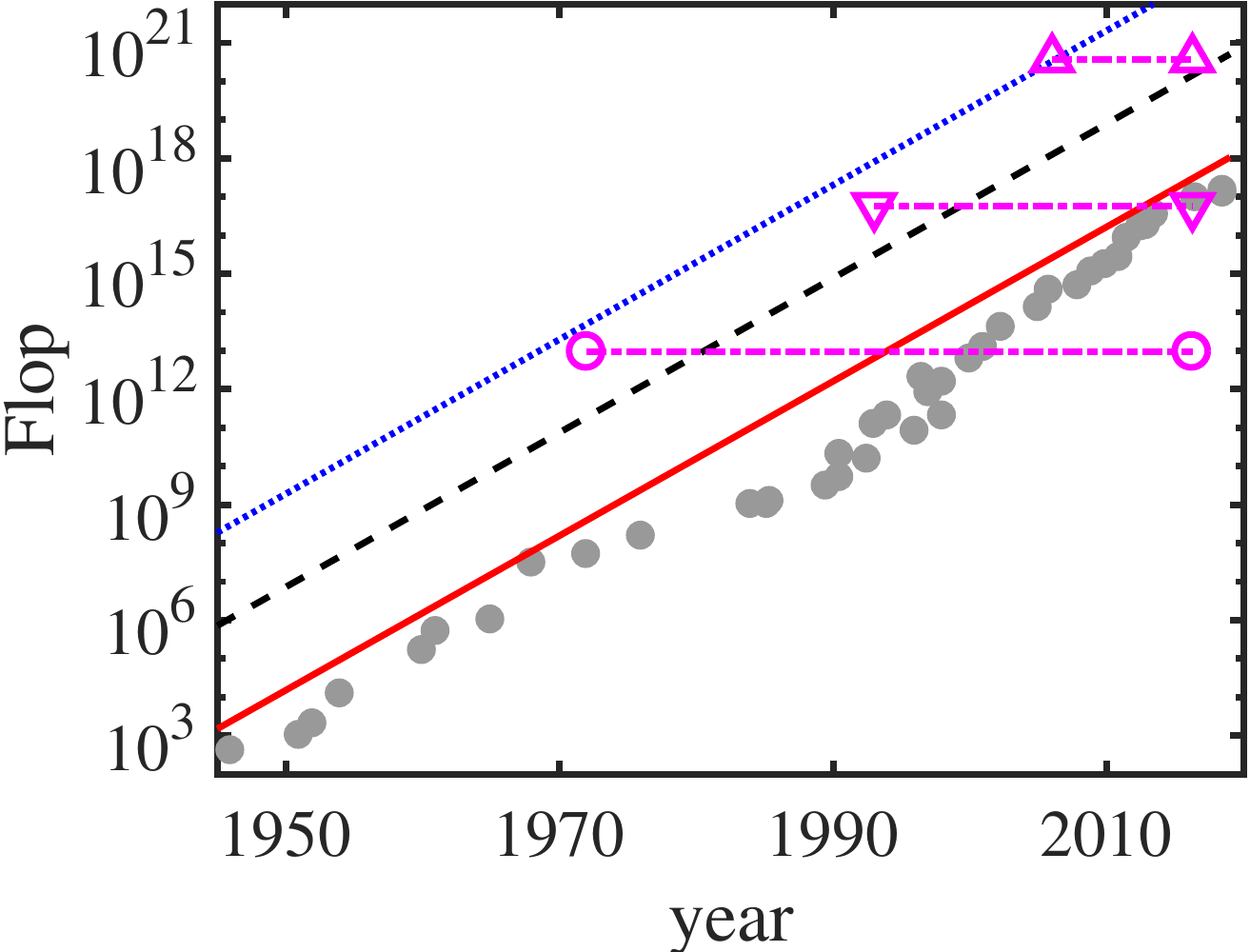}}%
\mylab{-.23\textwidth}{.37\textwidth}{(a)}%
\hspace*{5mm}%
\raisebox{0mm}{\includegraphics[height=.35\textwidth,clip]%
{\figpath 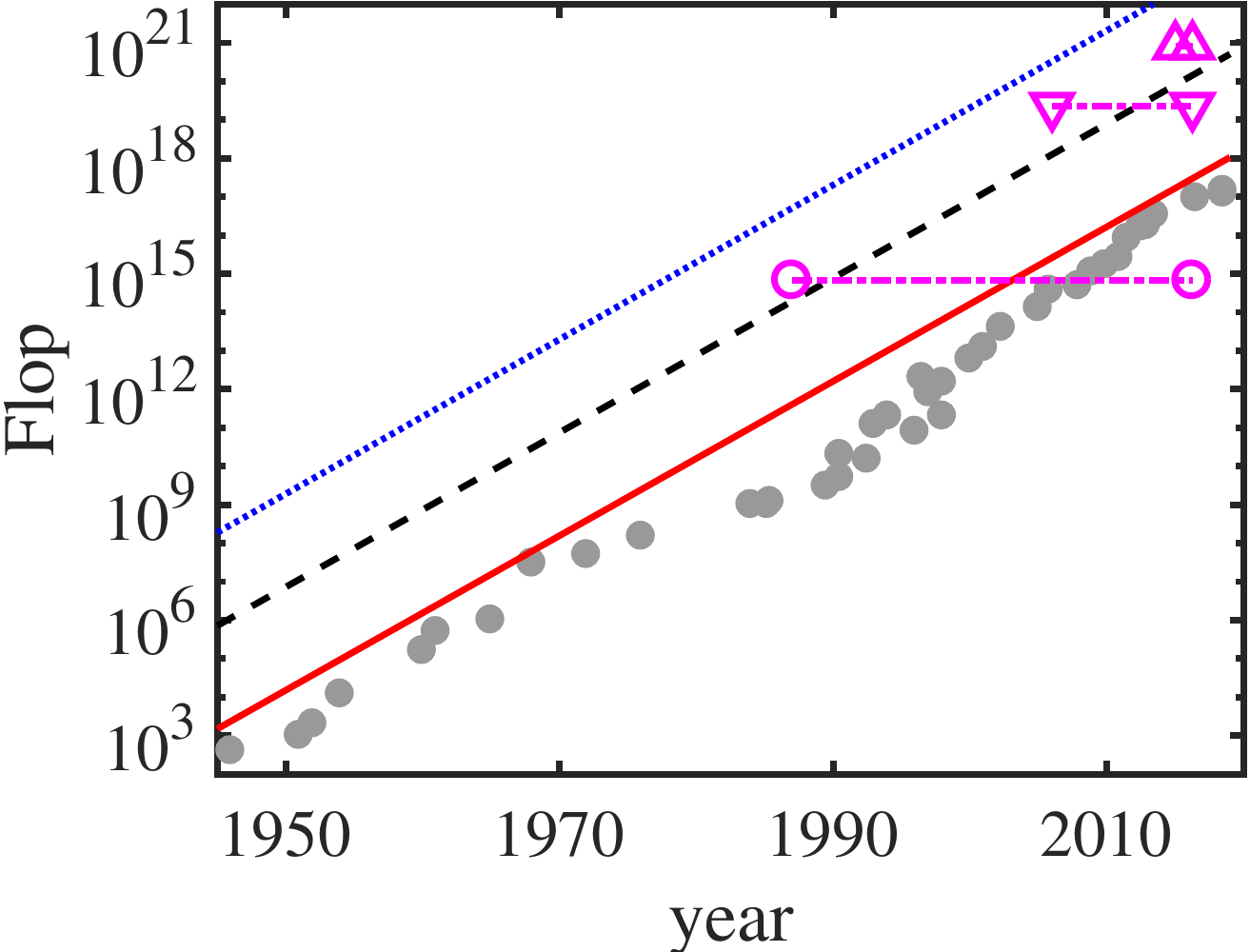}}%
\mylab{-.18\textwidth}{.37\textwidth}{(b)}%
\caption{%
(a) The grey closed symbols are the speed of the fastest computer as a function of the year
of commission, in floating-point operations (flop) per second. Trend lines are: \solid,
flops per minute; \dashed, overnight; \dotted, three months. They are reduced by a factor of
four with respect to nominal values to account for practical inefficiencies
\cite{gordonbell:17}, and grow by $1000$ every 15 years. The horizontal lines are the number
of operations required for simulations of isotropic turbulence, starting at the year of their
initial publication: \circle, $Re_\lambda=35$ \cite{orspat72}; \dtrian, $Re_\lambda=150$
\cite{jwsr}; \trian, $Re_\lambda=650$ \cite{kaneda06}.
(b) As in (a), for turbulent channels: \circle, $Re_\tau=180$ \cite{kmm}; \dtrian,
$Re_\tau=2000$ \cite{hoyas06}; \trian, $Re_\tau=5200$ \cite{lee:moser:15}.
}
\label{fig:flops}
\end{figure}

Estimating when a particular turbulent flow becomes computable depends on the details of the
flow and of the code, as well as on economic factors, but some approximate rules can be
derived from past experience. Consider the compilations in figure \ref{fig:flops}.
Simulations can be classified as `heroic', lasting for months, `routine', which can be run
overnight, and `trivial', which run in a few minutes. The first ones are research projects
that are typically performed only once, or at most a few times. The second ones are
industrially useful, and although still expensive when measured in terms of researcher's
time, they can be run several times if considered worthy, such as in testing well-developed
hypotheses. Trivial simulations can be run many times at little cost, and, as we will
discuss below in more detail, may be used to test or propose ideas. The trend lines in
figure \ref{fig:flops} show that computer speed increases by a factor of 1000 every fifteen
years, implying that a heroic job becomes routine in approximately twelve years, and trivial
in another twelve.

The cost of turbulence simulations depends primarily on their Reynolds number, which
determines the ratio between the largest and the smallest scales, and therefore the number
of grid points and operations required. For the simplest case of isotropic turbulence in a
triply periodic box, the relevant Reynolds number is $Re_\lambda=2K(5/3\diss\nu)^{1/2}$,
where $K$ and $\diss$ are, respectively, the kinetic energy and the dissipation per unit
volume, and $\nu$ is the kinematic viscosity. It can be shown that the linear range of
scales is proportional to $Re_\lambda^{3/2}$ \cite{tenn}, so that the number of operations
is proportional to $(Re_\lambda^{3/2})^4=Re_\lambda^6$ (because there are three spatial
dimensions plus time). The first simulation of this flow appeared in 1972 \cite{orspat72},
at the marginally turbulent $Re_\lambda=35$ in a box of $32^3\approx 32,000$ Fourier modes.
It is represented in figure \ref{fig:flops}(a) by the horizontal line with circles starting
at the year of its first publication. The other two horizontal lines correspond to later
simulations at higher Reynolds number, and it is clear from the figure that each of these
simulations was heroic at the time. It is also clear that the first two simulations have by
now become trivial, and that the largest one is becoming routine. In fact,
$Re_\lambda\approx 1000$ is becoming commonplace, and some $Re_\lambda\approx 2000$
simulations have been reported at somewhat reduced resolution and shorter evolution times.

More complex flows are more expensive to simulate. Figure \ref{fig:flops}(b) displays the
operation count and publication year for turbulent channels, which are inhomogeneous and
anisotropic because of the presence of the wall. The relevant Reynolds number is $Re_\tau
=u_\tau h/\nu \sim Re_\lambda^2$, where $u_\tau$ is the friction velocity and $h$ is the
channel half-height. The number of operations scales like $Re_\tau^3
Re_\tau^{3/4}=Re_\tau^{15/4}$ and, because the anisotropy requires larger computational
boxes than the isotropic case, and the numerics are somewhat harder, the first simulation
only appeared in 1987 \cite{kmm}, for $Re_\tau=180\, (Re_\lambda\approx 60)$ using $4\times
10^6$ grid points. That case became trivial several years ago. The second simulation in the
figure, $Re_\tau=2000$, is now routine, but the last one, $Re_\tau=5200$, will remain large
for some time.

Because the Navier--Stokes equations are generally believed to represent well the motion of
fluids, and the theory of numerical approximation is well developed, direct simulations are
essentially equivalent to experiments \cite{jim03jot}. Some limitations are
different in the two cases, but equivalent. For example, the boundary conditions in numerics
are equivalent to the details of tripping and of wind tunnel configuration in the
laboratory. The question of whether simulations should be pursued at higher Reynolds
numbers is thus equivalent to whether experiments should be. The main advantage of the
former with respect to the latter is that simulations have less observability limitations.
It is essentially impossible to simulate a flow without knowing everything about it. All
variables are required to integrate the equations of motion, or can be computed by
postprocessing. That is the reason for the high cost of simulations, but it gives them an
edge over experiments from the point of view of what can be learned. Even when a
high-Reynolds number experiment is performed, what can be measured is usually limited, and
the comparison between simulations and experiments should be made at equivalent levels of
observability. At present, heroic simulations are probably ahead of experiments in this
respect for simple flows, such as the ones mentioned above, and recent results from
both simulations \cite{lee:moser:15} and experiments \cite{ciclope:2019} suggest that it is
doubtful whether further increasing the Reynolds number would provide more than incremental
pay-offs. A rough estimate of the maximum required Reynolds number for channels, based on a
separation of scales of two orders of magnitude, is $Re_\tau=10^4$ \cite{jim12_arfm}, and
such a simulation is currently under way \cite{hoyas18}. For comparison, the Reynolds number
of a zero-pressure-gradient boundary layer over the dimensions of a flying bird ($L=0.25$~m,
$U=5$~m/s) is $Re_\tau \approx 200$, and the one over the fuselage of a medium-sized
airliner ($L=25$~m, $U=300$~m/s) is $Re_\tau \approx 40000$. 

A quick survey of recent large turbulence simulations shows an increase in the
number of more complex flows, either from the point of view of geometry \cite{gun:mac:sim:sor:16,Vinuesa:etal:18}, or of rheology \cite{Teng:etal:18}, in
detriment of simpler canonical ones.

Two caveats should be mentioned regarding this comparison between simulations and
experiments. The first one is that the fuller information from simulations is not always an
advantage, because it may not be necessary to know everything about the flow for some
particular purpose. If what is required is to answer a specific question, it may be enough
to perform an ad-hoc simulation or experiment to answer it, even if other variables are not
obtained. These are what we have defined above as `routine' simulations (or experiments), and
which technique to choose in each case should be guided by economy and need.

The second caveat is that operation counts may be starting to lose importance as the chief
constraint for simulations. Large computations generate huge amounts of data, which, as we
will see in the next section, may take a long time to postprocess. Individual flow fields of
the largest simulations in figure \ref{fig:flops} have file sizes of the order of
Terabytes. Minimal statistical significance requires at least 100 such fields, and full
advantage of a simulation can only be gained from time-resolved sequences of several
thousand files. The resulting multi-Petabyte data sets are not only expensive to store and
maintain, but hard to share among interested groups \cite{pha_jhu,sillero:iop16}, although
sharing is one of the main reasons to compile them in the first place. It is becoming
increasingly common to process data on the fly \cite{lozano-time}, without keeping most of
them, or to run simulations at full resolution while storing only underesolved data
\cite{encinar18}. But doing so detracts from their value. The storage and sharing problem is
not restricted to simulations. It is related to the range of scales in high-Reynolds number
flows, and is shared by any experiment able to retrieve relatively complete information
about a flow \cite{butl:96}.

Finally, it should be clear that neither simulations nor experiments should be confused with
theoretical understanding. Simulations, by providing detailed flow information, and by
catalysing experiments to do the same, is changing our views of turbulence in the same way
as the hot wire, the Pitot tube or flow visualisation did before. But taking advantage of
these opportunities requires a separate set of activities, which are addressed in the next
section.

\section{Overnight simulations and conceptual turbulence}\la{sec:structures}

It is a welcome characteristic of fluid mechanics that we know the equations of motion, and
that we can simulate the subject of our study to any desired precision. Moreover, we saw in
the last section that computers have evolved to the point at which simulations can address
what is probably the asymptotically relevant set of parameters in simple flows. It follows
that, once computers become fast enough for these simulations to become routine, they can
provide answers to any concrete question about any specific flow, and that, in that sense,
`we know everything' about it.

This knowledge can take several forms. The result of a simulation is a data
set, which can be a single number, a collection of space- and time-resolved flow fields, or
anything in between.

Consider first the case of a single number, such as the K\'arm\'an constant of the
logarithmic law of the wall, $\kappa\approx 0.4$ \cite{Hultma12}, or the prefactor of the
Kolmogorov energy spectrum, $C_K\approx 1.6$ \cite{sreeni95}. These values were originally
obtained experimentally but, if we simulate turbulence at high enough Reynolds number, the
results are very close to the experiments, and even suggest corrections
\cite{kaneda03,lee:moser:15}. Moreover, if we repeat the simulations several times, they
always return the same result, and, in this sense, it can be argued that the machine `knows'
the value of the constants, and `needs' no further clarification. It is up to us to decide
whether this should be interpreted to mean that the machine understands `why' the value of
$\kappa$ is 0.4, as opposed to just knowing that it is so.

The question of `understanding' is probably impossible to answer in any general sense
\cite{Nagel:74}, but there are phenomenological definitions that may help within the
narrower framework of computable physical processes. For example, we may insist in having a
simple way of estimating the approximate result of experiments, such as why $\kappa$ is not
$10^{-2}$ or $10^3$, or we may want to know how to manipulate the flow so that 
$\kappa=0.8$, or $\kappa=0.2$.

None of these questions can be easily answered by `natural' simulations or experiments. But
they are often accessible to conceptual `thought' experiments, which may or may not be
physically realisable. Such experiments have a long history in physics, and in some cases
require little more than pencil, paper, and imagination. In turbulence, where the outcome of
a given modification is often hard to guess, they usually also have to rely on simulations.

The first conceptual computer experiments in turbulence were probably those intended to
clarify whether the statistics of isotropic turbulence, an intrinsically dissipative
phenomenon, depend on the particular dissipation mechanism of regular viscosity. In a sense,
the question had already been answered by large-eddy simulations (LES), which rely on
dissipation models very different from viscosity, and which were known to provide accurate
answers even in the absence of adjustable parameters \cite{rog:moi:84,germetal91}. Model
simulations using unphysical viscosity \cite{borue95}, or even no viscosity at all
\cite{she93}, soon confirmed that the statistics of the inertial range of scales are
independent of the dissipation details.

Conceptual experiments on wall bounded turbulence appeared at approximately the same time.
Also in that case, LES reproduces the statistics of the logarithmic and
near-wall layers \cite{kim:83}, but the related questions about the nature of the
interactions between the wall and the flow \cite{jpin}, the interplay between the inner and
outer layers \cite{jpin,miz:jim:13}, or the importance of large-scale chaos as opposed to
individual structures \cite{jmoin,oscar10_log}, required unphysical simulations that would
have been difficult to reproduce experimentally.

These conceptual experiments were made possible because the required simulations could be
performed in at most a few days, but another type of data sets are useful even when their
generation and analysis tends to stretch over months or years. We have already mentioned
that an essential difference between simulations and experiments is that the former have no
observational problems. Everything can be measured and everything can be stored, permitting
iterative interrogation of the data. The results can be processed, conclusions drawn, new
questions posed, and the data re-analysed to answer them.

\begin{figure}[t]
\centering
\begin{minipage}[t]{.47\textwidth}%
\includegraphics[width=.95\textwidth,clip]{\figpath 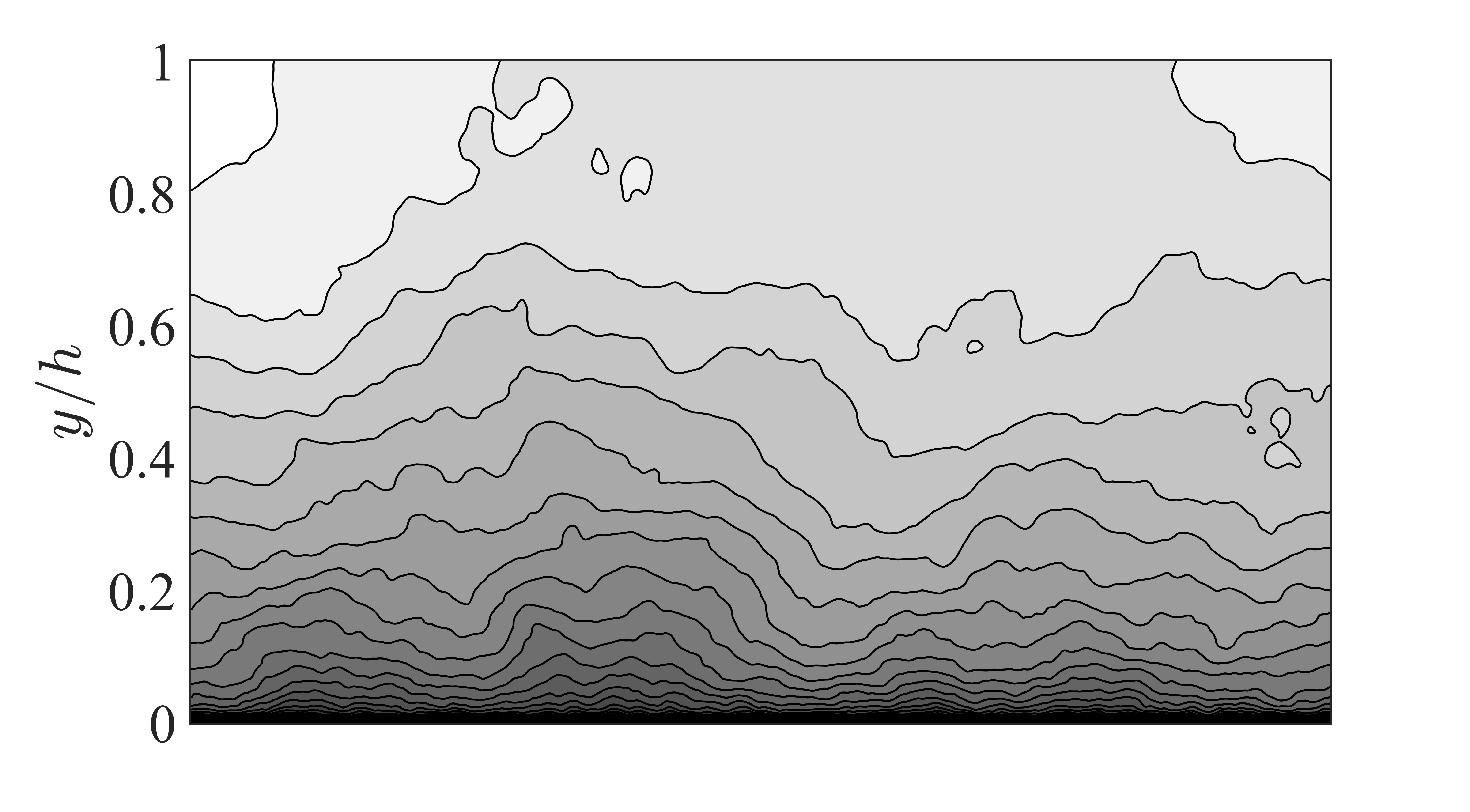}%
\mylab{-1.03\textwidth}{.35\textwidth}{(a)}%
\\[-5mm]%
\includegraphics[width=.95\textwidth,clip]{\figpath 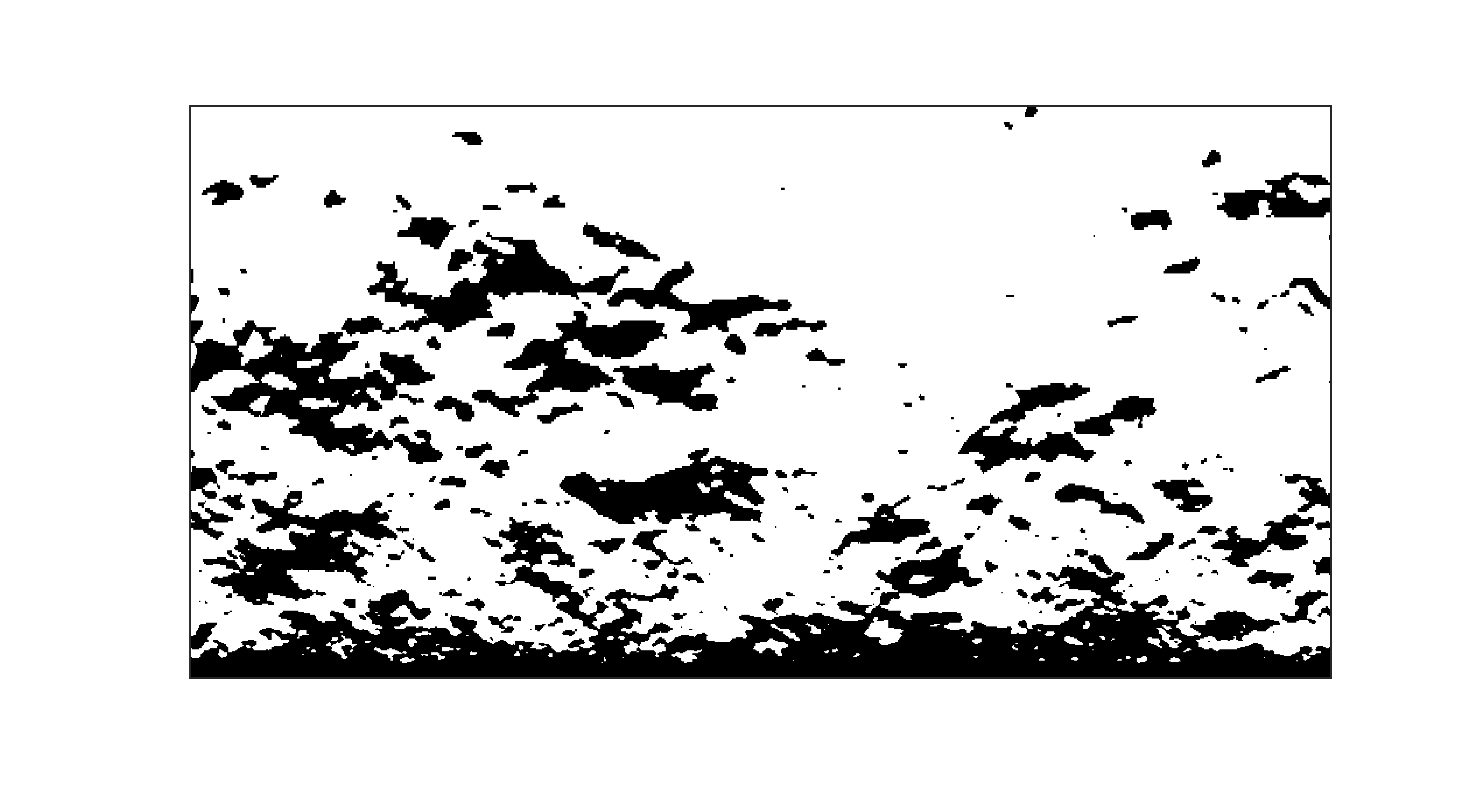}%
\mylab{-1.03\textwidth}{.35\textwidth}{(c)}%
\end{minipage}%
\hspace*{-7mm}%
\begin{minipage}[t]{.47\textwidth}%
\includegraphics[width=1.01\textwidth,clip]{\figpath 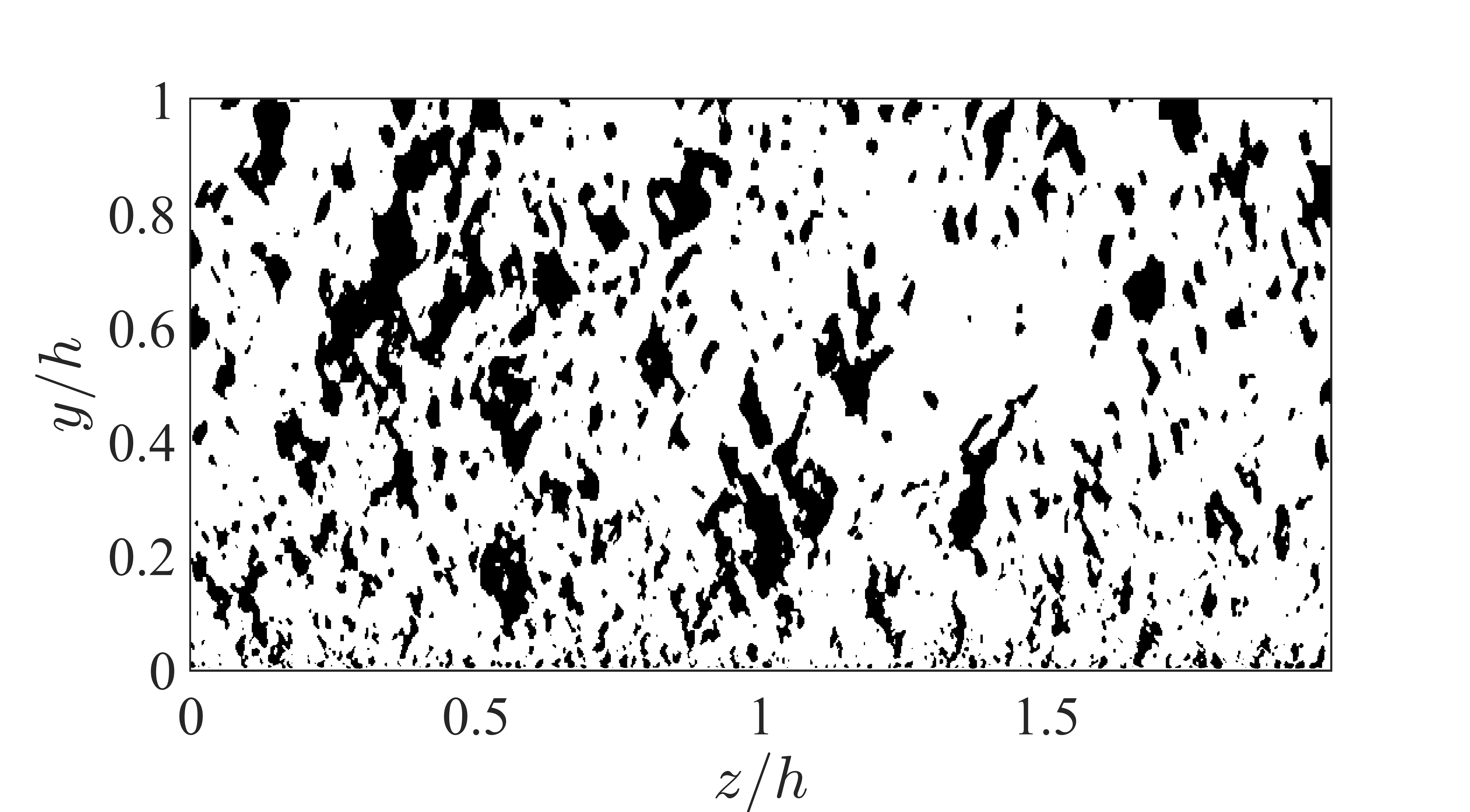}%
\mylab{-.05\textwidth}{.35\textwidth}{(b)}%
\\[-5.5mm]%
\includegraphics[width=1\textwidth,clip]{\figpath 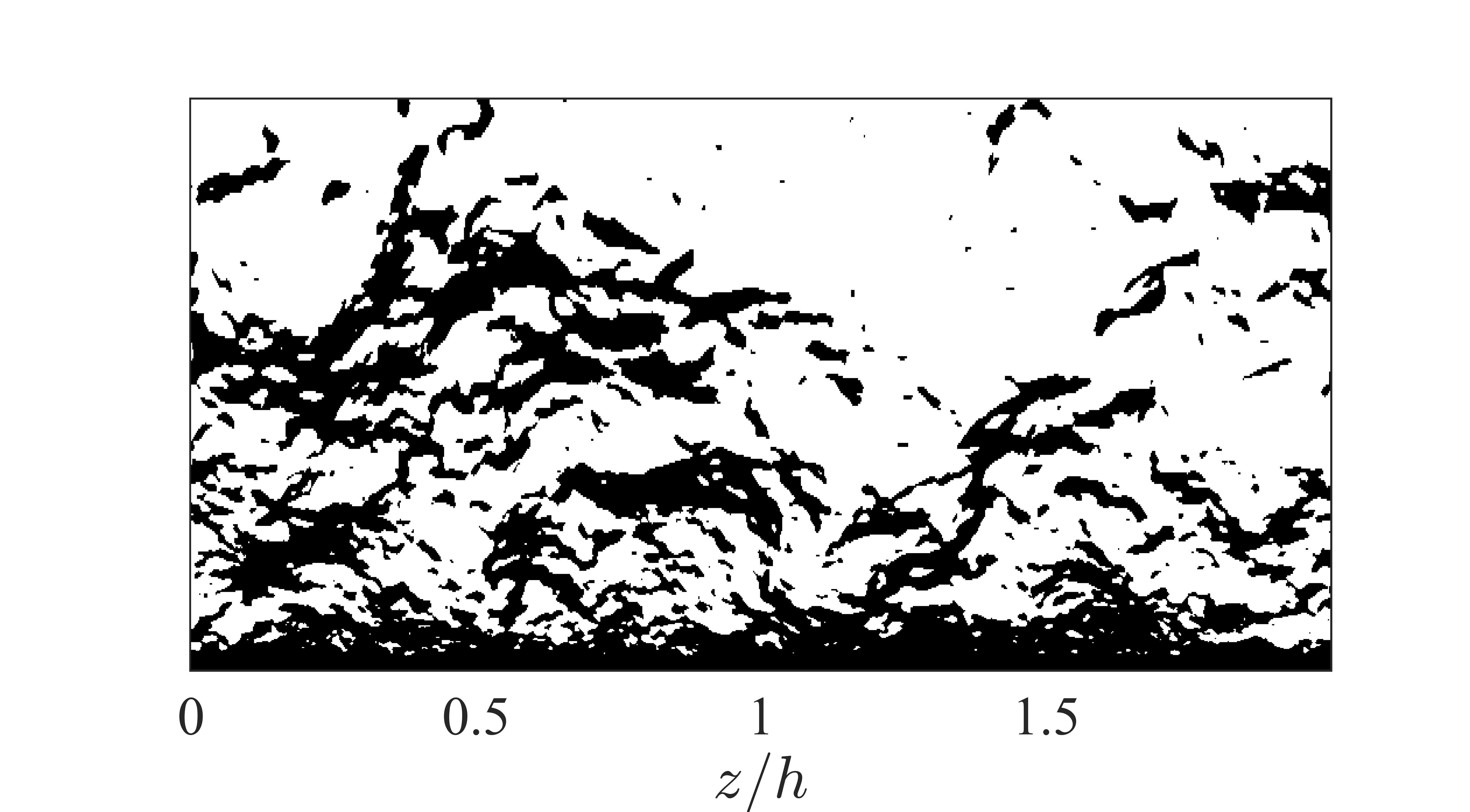}%
\mylab{-.05\textwidth}{.32\textwidth}{(d)}%
\end{minipage}\\%
\centerline{\includegraphics[width=.65\textwidth,clip]{\figpath 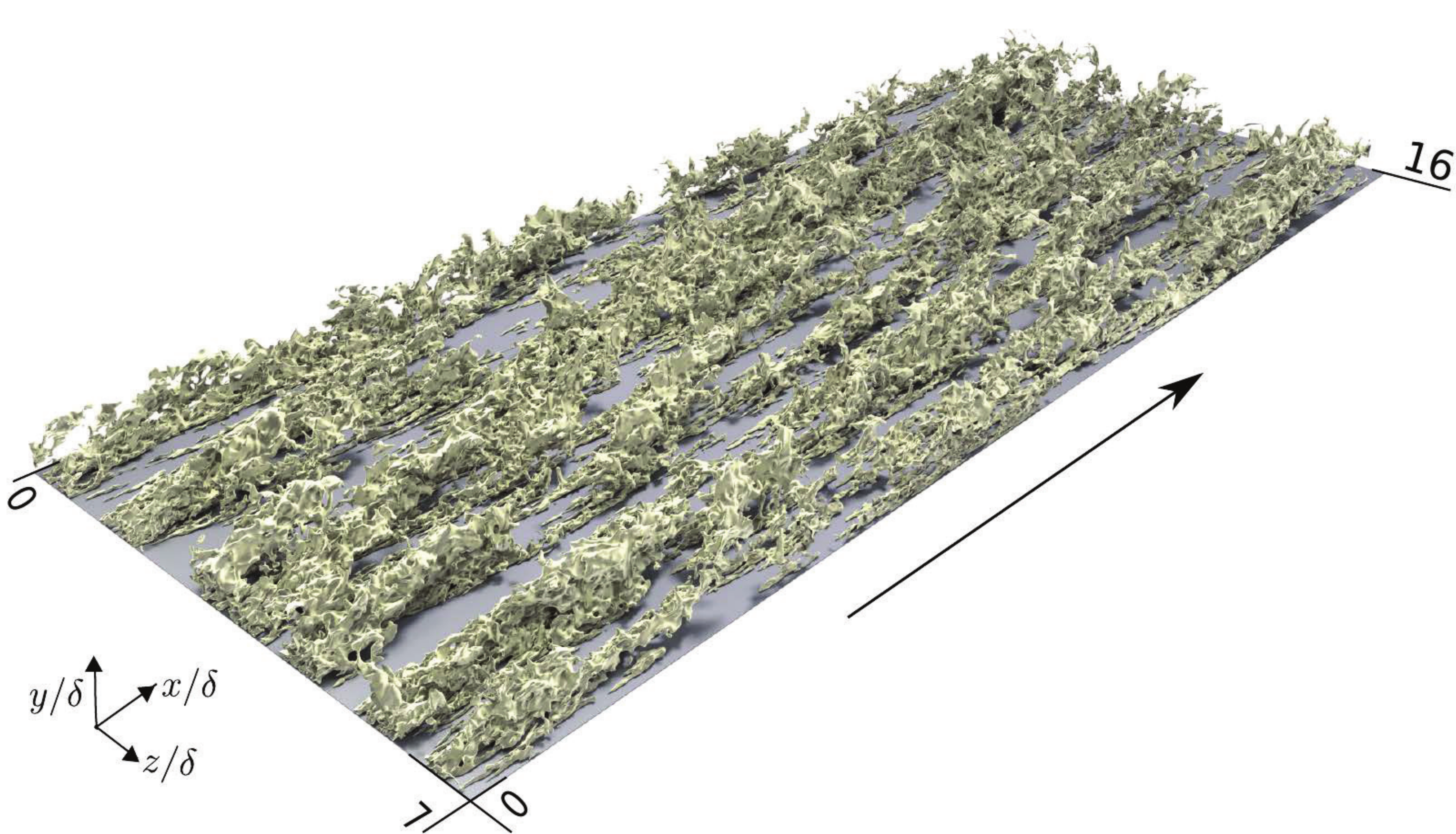}%
\mylab{-.55\textwidth}{.27\textwidth}{(e)}%
}
\caption{%
(a) Cross section of the streamwise velocity, $U$, in a channel at $Re_\tau=2000$
\cite{hoyas06}. Contours are $U/\utau=$(0:\,0.78:\,25), increasing upwards.
(b) Vertical derivative of the flow in (a). Dark areas are $\p_y U >3 S$ where $S=\bra \p_y
U\ket_{xz}$ is the plane-averaged mean shear.
(c) Spanwise derivative. $\p_z U >1.5S$. 
(d) Magnitude of the cross-plane gradient. $[(\p_z U)^2+ (\p_y U)^2]^{1/2} >3S$. 
(e) Outer-flow high-velocity streaks, visualised by the perturbation streamwise velocity,
$u=2\utau$. Boundary layer at $\retau\approx 1200$--1400. Structures shorter than half the
boundary layer thickness, $\delta$, have been removed (reproduced with permission from
\cite{jim18}; credits J.A. Sillero)
}
\label{fig:grads}
\end{figure}

This is not necessarily cheap, even if processing a single flow field is usually equivalent
to a single time step of the simulation. We have already alluded at the end of the previous
section to the difficulty of storing large data sets. Using them is also computationally
expensive, because many snapshots have to be analysed, typically iteratively. A rough
estimate from the experience of our group is that postprocessing the data set resulting from
a simulation requires from two to three times more computer time than the original run. But
postprocessing can (and should) be done at leisure over several years, and in collaboration
with teams beyond the original simulators, and it provides opportunities that would
be unavailable in any other way.

A simple but striking example of how simulation data can help correct misinterpretations due
to instrumental limitations is figure \ref{fig:grads}(a-d), which shows a cross-stream
section of the streamwise velocity in a channel, and three versions of the `vorticity
clefts' detected by different implementations of the cross-plane gradient. The
one-directional gradients in figure \ref{fig:grads}(b), which simulate the two-dimensional
sections in a particular experiment \cite{meiadr95,adrmeintom00}, could naively be
interpreted as planar fronts separating two-dimensional uniform-momentum layers, and this
interpretation has occasionally persisted \cite{klewicki07} even if the original group
specifically warned otherwise and quantitatively characterised the three dimensional
structure \cite{tomadr03}. The two-dimensional gradients in figure \ref{fig:grads}(d)
clearly show the tube-like geometry of the uniform-velocity streaks, as is also seen in the
three-dimensional representation in figure \ref{fig:grads}(e). The vertical clefts in figure
\ref{fig:grads}(c) show how easy it is to introduce artefacts from incomplete data.

Another illustrative case is the characterisation of sweeps and ejections in wall-bounded
turbulence, which had originally been defined from single-point hot-wire data
\cite{wall:eck:bro:72,lu:wil:73}. The argument was that, because the tangential Reynolds
stress can be written as the average of the product of the streamwise and wall-normal
velocity fluctuations, $-\bra u v\ket_{xz}$, where $\bra\ket_{xz}$ is the average over
wall-parallel planes, the regions where both velocities are strong and either $u>0$ and
$v<0$ (sweeps) or vice versa (ejections) mark significant locations for momentum transfer.
In practice, these two types of regions carry about 60\% of the total tangential Reynolds
stress in attached wall-bounded turbulent flows \cite{jim18}.

A lot of work has been devoted to such `quadrant' analysis, and a lot has been learned about
it from hot wires and particle-image velocimetry \cite{chradr01}. This helped to
establish the `hairpin packet' model of wall turbulence \cite{adr07}, although the limited
information provided by the experiments, usually at most restricted to two-dimensional
sections, resulted in some confusion about the geometrical properties of the objects in
question.

It was only when full three-dimensional vorticity and velocity fields became available from
direct simulations of turbulent channels and boundary layers that the true geometry of the
sweeps and ejections could be clarified \cite{robinson91,jc06_vor,lozano-Q}. They are
three-dimensional structures which separate into wall-attached and wall-detached
families, depending of whether their root reaches the wall or not. Most of the momentum
transfer is associated with the wall-attached structures, which form a self-similar family of
eddies with moderate aspect ratios (length:\,width:\,height $\approx$ 3:1:1), organised as
spanwise pairs of one sweep and one ejection. These pairs can
be interpreted as the two sides of approximately streamwise rollers, which, in turn, are
organised as rows along the edges of the streamwise velocity streaks. Sweeps, bringing
high-speed fluid towards the wall, are located in the high-velocity streaks, and ejections,
moving low-speed fluid away from the wall, are located in the low-speed ones.

These composite objects come in all sizes, from packets of individual vortices near the
wall, with spanwise width of the order of 100 wall units and streamwise length of 500--1000
wall units \cite{robinson91}, to very large structures of the order of the boundary layer
thickness, formed by vortex tangles instead of by single vortices
\cite{jc06_vor,clus_osc,lozano-Q}.

Given the importance of wall-attached structures, it is a fair question whether they form at
the wall and rise, appear away from the wall and sink, or any combination of the
two. The resolution of this question either requires the conceptual simulations discussed
above, or tracking the structures along their full lifetime. The latter is difficult in the
laboratory, because structures travel long distances from the moment they first appear to
the point when they can be recognised as strong events, and also because they drift in and
out of the two-dimensional sections typically provided by experiments. The answer came
from simulation data sets in which the flow variables are stored closely enough in time for
individual structures to be followed \cite{lozano-time}. It turns out that sweeps are formed
away from the wall and move down, and that ejections are formed near the wall and move up.
The objects formed by the sweep-ejection pairs do not migrate significantly up or down.

The lack of vertical drift of the pairs suggests that their cause is not the wall but the
shear, which, in any case, is the ultimate source of kinetic energy for the turbulent
fluctuations \cite{tenn}. This was confirmed in \cite{dong17}, who repeated the
three-dimensional structural analysis for a uniform shear flow with no walls. The results
were similar to those in channels, although with some interesting differences. Sweeps and
ejections are found in both flows, with similar aspect ratios, and they are located in both
cases at the interface between a high- and a low-velocity streak. But the inclination of the
associated rollers is different in the two flows. The average inclination in a uniform shear is
45\degree, as required by symmetry considerations, but it becomes shallower for
detached structures in channels, and almost parallel to the wall for attached pairs. At the
same time, the ejection, which is typically located below the sweep, because it comes from
the lower-speed fluid below, is of the same size as the sweep in the uniform shear, but
shrinks as the pair gets closer to the wall. The same is true for the low-speed streak. The
overall conclusion is that the effect of the wall is to hinder the formation of the inclined
rollers and of the streaks, rather than to promote them.

Consistent with their location in high- and low-velocity streaks, sweeps advect faster than
ejections, and the initially spanwise pair slowly separates in the streamwise direction at
the same time as it does vertically. The velocity difference in the two directions is of the
order of the friction velocity, $\utau$, so that the lifetime of the roller can at most be
of the order of its turnover time $O(y/\utau)$, where $y$ is both the distance from the wall
and the height of the attached eddies. This was confirmed by direct tracking in
\cite{lozano-time}.

It should be clear from the previous discussion that the importance of archived data sets is
mainly that they can be used to test theories. Consider a final example. It was known that
the main contributor to the evolution equations in shear flows is advection by the mean
profile \cite{jim13_lin}, and we have already mentioned that the source of kinetic energy
for the fluctuations is their interaction with the mean shear. In that sense, the
energy-producing scales of shear turbulence are linear, although nonlinearity dominates
further down the cascade. It was established in \cite{corrsin58,sadd94} that the criterion
separating the two regimes is the ratio between the eddy turnover time and the deformation
time defined by the inverse of the mean shear. When this Corrsin number is large, the shear
of the mean velocity profile dominates, and the flow can be considered linearisable. Using
only spectral information, the Corrsin number can be defined for the mean flow, or
individually for eddies of a given size. The results is that wall-bounded flows are
linearisable in the logarithmic and buffer layers, and that the linearisation can only be
applied to eddies which are at least as large as their distance from the wall \cite{jim18}.
Geometrically, this suggests that the attached structures, defined above as reaching the
wall, are identical to the `active' eddies hypothesised in \cite{Townsend1976}, and that both
are just eddies larger than the spectral Corrsin scale. This was tested in \cite{dong17} using
the uniform shear flow. When discussing this flow, we
mentioned that the sweeps and ejections in the uniform shear were essentially the same as
the attached structures in channels, but this begs the question of which eddies are to be
considered attached in a flow without walls. What \cite{dong17} showed is that the
uniform-shear eddies equivalent to the attached ones in channels are those larger than the
Corrsin limit. Smaller ones behave like the detached channel structures.

Many similar examples can be mentioned of the use of computers in supporting or falsifying
theories or models, from mostly linear ones \cite{ala:jim:06,McKSha2010,jimenez:2015} to the
fully nonlinear identification of invariant solutions \cite{KawEtal12}. Discussing them any
further would take us beyond the scope of the present paper, but a recent survey of the
properties of structures in wall turbulence and related flows is \cite{jim18}.

\section{Trivial simulations and Monte Carlo research}\la{sec:curio}

In the previous section we have given examples of how using computers as `universal
answering' machines, performing routine searches and conceptual experiments in no more
than a few hours, can advance the study of turbulence. The procedure follows 
the classical scientific method of designing experiments to address a particular
question, with the only difference that they are performed computationally. However,
because the time spent in each simulation is short, but not trivial, testing more than a few
possibilities still requires a `heroic' multi-month effort. A consequence is that
experiments tend to be designed carefully, because wild guesses waste valuable simulation
time, and that they tend to address incremental questions, rather than revolutionary ones.

A different approach becomes possible once conceptual experiments are cheap enough to be
executed at essentially no cost. It is then possible to perform `silly' experiments, just in
case they produce something, and to decide a posteriori which results are interesting. In
essence, this is equivalent to using a Monte Carlo method to compute the volume of a
hypersphere by randomly throwing points and counting which fraction falls within the sphere
\cite{Hammer:64}, instead of by careful integration.

Once this level of computational performance has been reached, simulations can be considered
tools for creating (or suggesting) questions, rather than just for generating
answers. For example, consider a machine that can add ordered pairs of integers by consecutively accumulating units (e.g. $2=1,1\Rightarrow 2+3=1,1,1,1,1$), but which
does not know the properties of addition. We can use it to test commutativity by asking
whether pairs such as $1+4$ and $4+1$ always give the same result, but this can only be done
if we suspect that commutativity exists.

A different approach would be to test a large number of random pairs looking for
repetitions, and to `notice' that several pairs give the same result: e.g., $1+9$, $2+8$,
etc. Once equivalent pairs are collected into classes (`10' in the previous example),
somebody, either the machine or the programmer, may look for possible regularities within
each class, and wonder, for example, why equivalent pairs tend to come in couples,
$2+8=8+2$. It is obvious that such an observation does not constitute a proof of
commutativity, but it may be considered as a suggestion that such a property exist, and an
invitation to research it more carefully. Although the example is trivial, it illustrates
the potential of randomised experiments to be statistically exhaustive, in the sense that
they can investigate properties with few preconceived ideas.

\begin{figure}[t]
\centering
\raisebox{0mm}{\includegraphics[height=.36\textwidth,clip]%
{\figpath 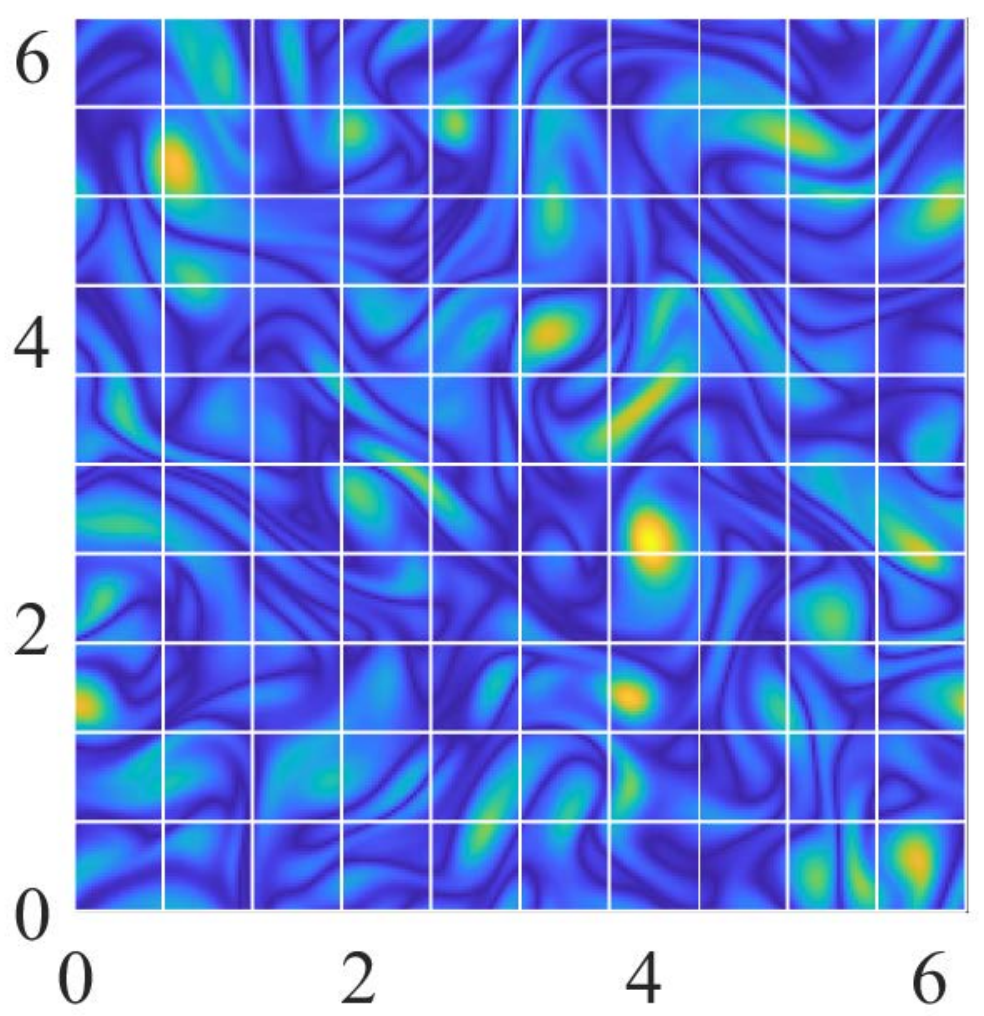}}%
\mylab{-.175\textwidth}{.38\textwidth}{(a)}%
\hspace*{3mm}%
\raisebox{0mm}{\includegraphics[height=.36\textwidth,clip]%
{\figpath 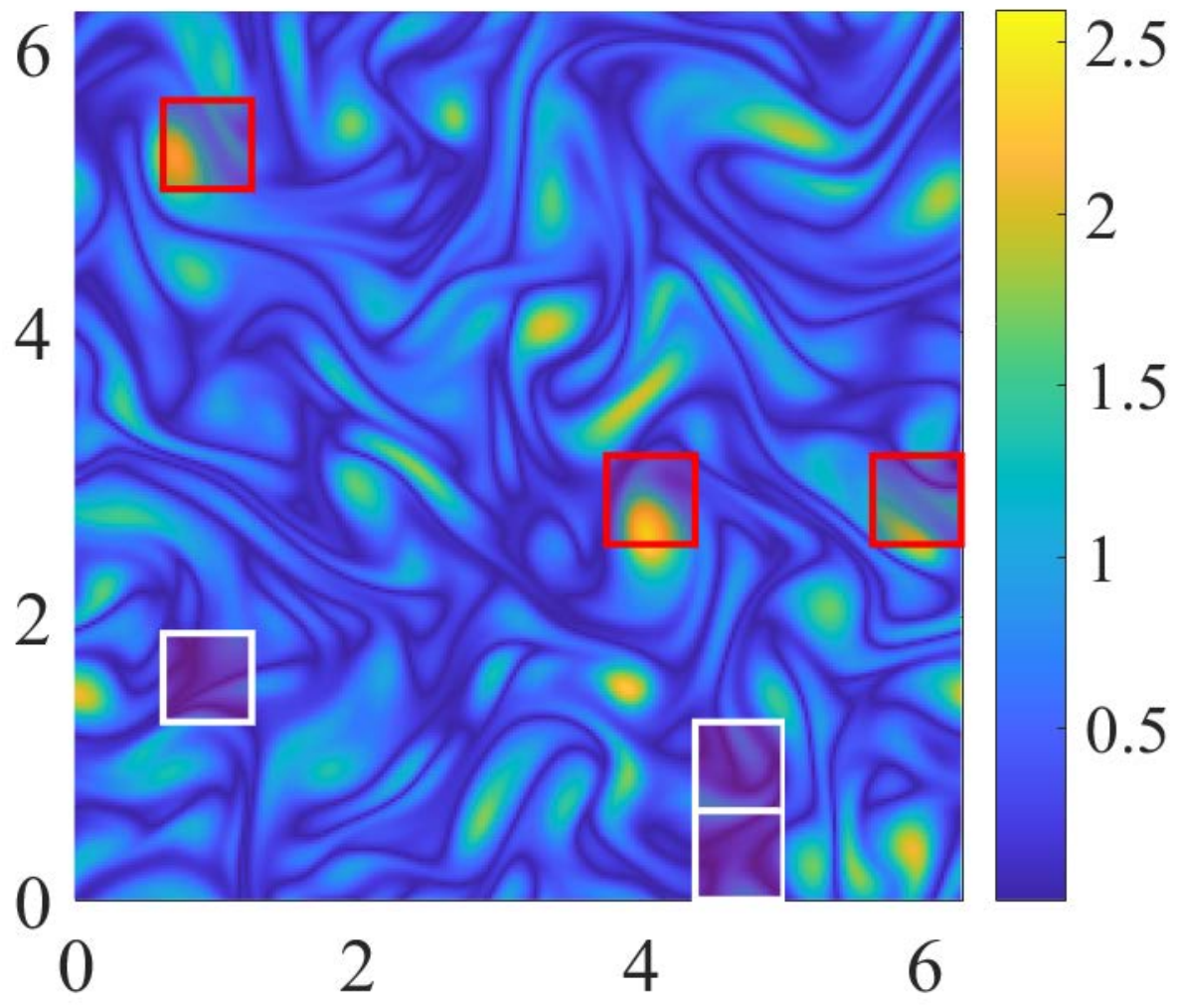}}%
\mylab{-.26\textwidth}{.38\textwidth}{(b)}%
\caption{%
(a) Vorticity magnitude in an initial condition to be tested. The overlaid grid is
used to select the test subsets.
(b) Cells outlined in white are the three least significant ones after $\omega'T=7$, where
$\omega'$ is the root-mean-square vorticity of the initial field. Those outlined in red are
the three most significant ones. In this experiment, the vorticity in the subset being
modified is substituted by a constant whose magnitude conserves the overall enstrophy, with
opposite sign to its original mean.
}
\label{fig:tur2d}
%
%
\vspace*{2mm}%
\centering\small
\begin{tabular}{c}
\hline
\begin{minipage}[t]{.90\textwidth}
\noindent{}%
\begin{list}{}{\setlength{\parsep}{0ex}\setlength{\itemsep}{0ex}\setlength{\topsep}{0ex}%
\setlength{\labelwidth}{0.5em}}%
\item[{\bf ProperLabelling$(g,T,A,\mathcal{N})$}] 
\item Given an initial flow field, $g(\vec{x},t=0)$,  a target time, $t=T$, 
\item and a partition, $A$, of $g$  into subsets.\vspace{1ex}
\item[]\la{ploff:2} {\bf For all} subsets $a\in A$
\item[]\la{ploff:20} \qquad {\bf Create} a test field by $\tg(a,t=0)\Leftarrow \mathcal{N}(g)$.
\item[]\la{ploff:3} \qquad {\bf Evolve}  $g$ and $\tg$ to $t=T$, using the equations of motion.
\item[]\la{ploff:30} \qquad {\bf Compute} the perturbation growth, $\diss=\|\tg(T)-g(T)\|$ 
\item[]\la{ploff:4} {\bf Iterate} over $A$.\vspace{1ex}
\item[\bf Output: ]\la{ploff:40} 
\item {\bf Label} as most (least) significant the $a$'s maximising (minimising) $\diss$.
\item[] {\bf Store} $a, \diss$ and all properties of the most (least) significant $a$'s.
\end{list}\vspace{1ex}
\end{minipage}\\
\hline
\end{tabular}

\caption{Labelling algorithm for selecting the most significant subsets of a flow $g$ under
the transformation $\mathcal{N}$.}
\la{alg:PLOFF}
\end{figure}
%
\begin{figure}[t]
%
\centerline{%
\raisebox{0mm}[0mm][0mm]{%
\includegraphics[height=.30\textwidth,clip]{\figpath 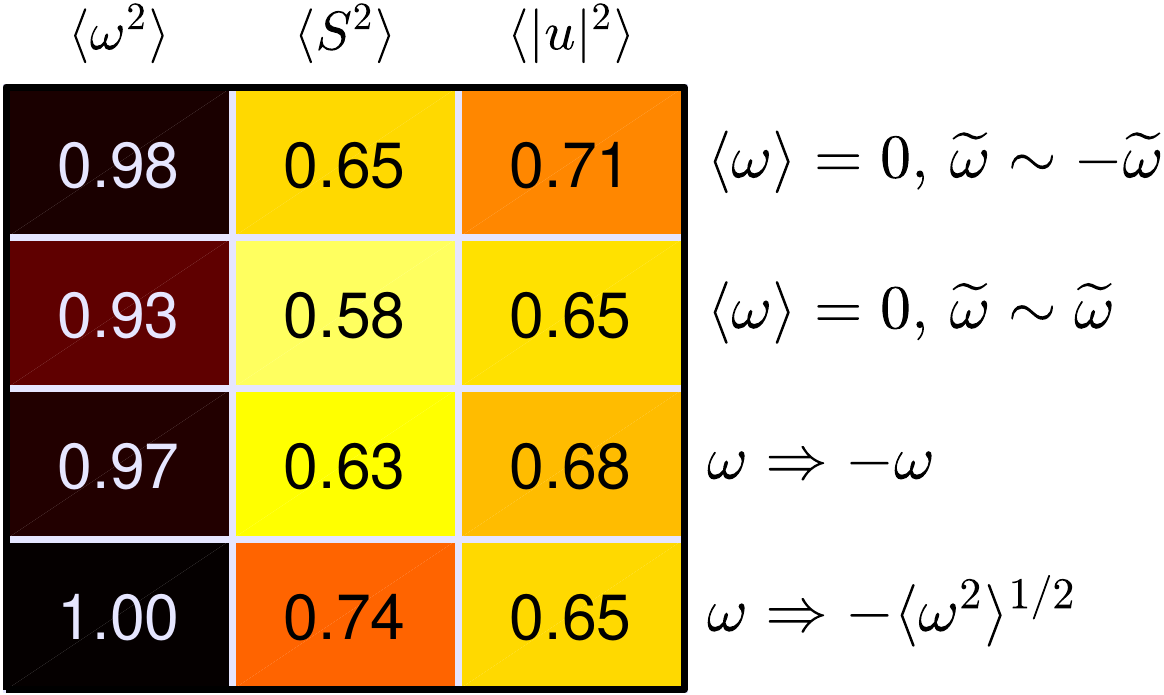}%
\mylab{-.56\textwidth}{.20\textwidth}{(a)}}%
\hspace*{3mm}%
\raisebox{-20mm}{%
\begin{minipage}[b]{.30\textwidth}%
\centering
\includegraphics[height=.95\textwidth,clip]%
{\figpath 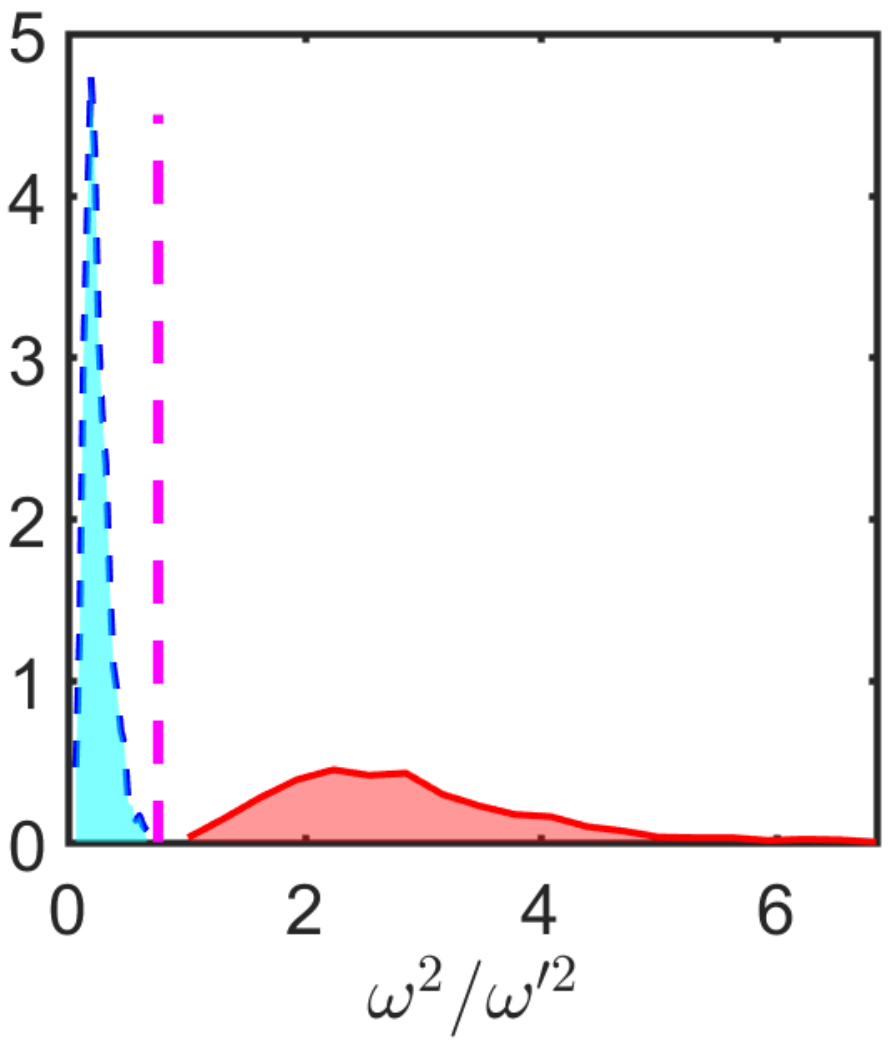}%
\mylab{-.19\textwidth}{.76\textwidth}{(b)}%
\vspace*{0mm}%
\\%
\hspace*{-1mm}\includegraphics[height=.95\textwidth,clip]%
{\figpath 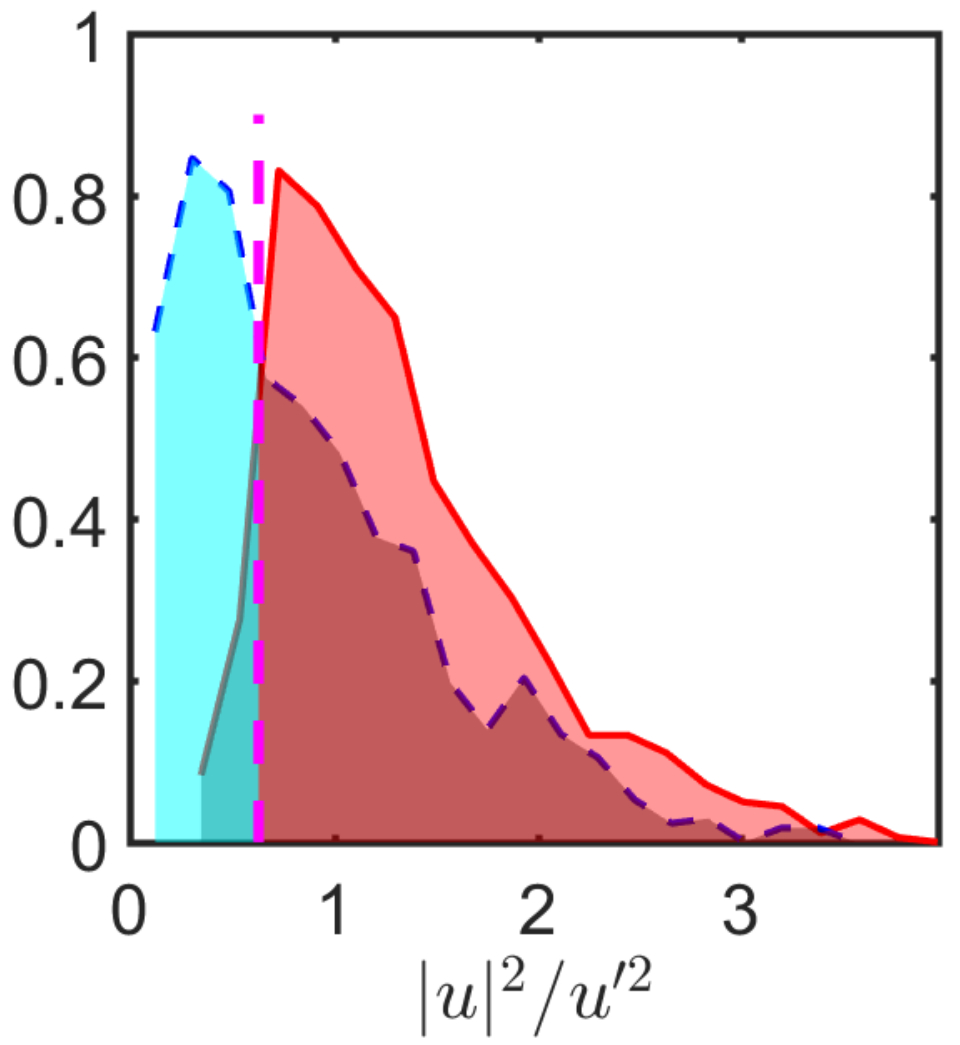}%
\mylab{-.19\textwidth}{.76\textwidth}{(c)}%
\end{minipage}
}}%
\caption{%
(a) Discrimination table for a set of experiments on the flow in figure \ref{fig:tur2d}. The
labels in the top row are the variables used for discrimination. Those on the right column
are the operations used to modify each cell, where $\bra\ket$ is the average over the cell,
and the tilde denotes fluctuations with respect to that average. Values in the table are
the discrimination efficiency illustrated in (b,c).
(b) Probability density function of the cell enstrophy for: \solid, most significant cells;
\dashed, least significant ones. The vertical line is the optimum discrimination threshold.
Conditions are as in figure \ref{fig:tur2d}, for the bottom line of the table in (a).
(c) As in (b), for the cell kinetic energy.
}
\label{fig:mynn}
\end{figure}

Consider how such a program could be extended to turbulence, using the example of the
identification of coherent structures, which we have already discussed in
\S\ref{sec:structures}, in relation with sweeps and ejections. The question is how to decide
whether there are regions in the flow which are more `significant' than others and, if they
are found to exist, to determine their properties. Coherent structures are not natural
constructs in fluid mechanics, because the flow is a continuous field with few obvious
boundaries along which it can be segmented. Structures have to be defined with some purpose
in mind, and tested carefully, because the human mind is prone to impose order even where
there is none. A cautionary example are the zodiac constellations, which have been
recognised by most cultures in one form or another, and imbued with predictive powers that
they are unlikely to have. The fact that most random binary predictions succeed 50\% of the
time has facilitated the survival of these spurious constructs, even in the absence of
supporting evidence, and it should give us pause that this performance is not much worse
than the observation, mentioned above, that the sweeps and ejections of wall-bounded flows
account for 60\% of the total tangential Reynolds stress.

Human researchers, who know the history of a subject, tend to look for incremental
improvements on that history, but machines are not biased in the same way. The ability of
intelligent machines to `reason' outside human prejudice, even when their intelligence is
limited, is probably the most interesting attribute of automatic experimental design.

A simple flow in which to test the automatic identification of structures is decaying
isotropic two-dimensional turbulence, where simulations are cheap enough to be run in
seconds. There is a wide consensus that this flow is controlled by the interaction of
compact vortices \cite{mcwilliams90b}, and that vortices should probably be found to be the
significant flow structures by any reasonable experiment. This `ground truth', together with its
relative accessibility to computation, makes two-dimensional turbulence a good example in which
to investigate some of the questions posed above. The following discussion follows closely the
analysis in \cite{jimploff18}.

Define significance as the ability of a modification of the initial conditions to change the future
behaviour of the flow. In particular, assume that we partition an initial flow
field, $g$, into a set of cells $a\in A$, and that we wish to test the effect of modifying
in some way the flow within one cell, $g(a, t=0) \Leftarrow \tg = \mathcal{N}(g)$ (see figure
\ref{fig:tur2d}a). The testing algorithm is sketched in figure \ref{alg:PLOFF}, and consists
of running the reference and modified flows, $g$ and $\tg$, to some target time $T$, and
measuring the growth of their separation, $\diss=\|\tg(T)-g(T)\|$. Changes that result in
a larger divergence for a given modification procedure are considered to have been applied
to more significant regions of the flow. The result is the identification of the (typically three
to five) most and least significant cells for each initial flow field and modification
procedure (see figure \ref{fig:tur2d}b). The final goal is to characterise which properties of a
cell can be used to discriminate whether it is significant or not, without running the
experiment again.

To that end, we collect as many properties as possible of the most (and least) significant
cells, such as the cell enstrophy, kinetic energy, etc., and repeat the process for
as many initial conditions (typically 100--200), partitions $(A)$, and experimental
transformations $(\mathcal{N})$, as desired. The result is a set of observations, classified
into disjoint significance classes, with which to train a  human or machine `researcher' to predict
significance in terms of cell properties, and to determine which cell properties are most
effective as discriminants in each case.

The example in figure \ref{fig:mynn}(a) shows the discrimination table for a set of
experiments performed on the flow in figure \ref{fig:tur2d}. The columns of the table are
possible discriminating variables, which in this case are the mean enstrophy,
$\langle\omega^2\rangle$, the mean rate of strain, $\langle S^2\rangle=\langle
2S_{ij}S_{ij}\rangle$, and the kinetic energy, $\langle |\uvec|^2\rangle=\langle
u_iu_i\rangle$, where $\bra\ket$ denotes averaging over the cell, and repeated indices imply
summation over $i=x,y$. The rows of the table are different modification strategies
$(\mathcal{N})$ for the initial conditions, all of which manipulate vorticity. For example,
in the bottom line, the vorticity in the cell is substituted by a constant, with opposite
sign to the original, and conserving the total enstrophy. The values in the table are the
discrimination efficiency illustrated in figure \ref{fig:mynn}(b,c). For each discriminating
variable, we construct the probability density function of the most and least significant
cells, and determine the threshold that best separates the two histograms. The
discrimination index is the fraction of cells that can be correctly classified in this way.
In well separated cases, such as the enstrophy in figure \ref{fig:mynn}(b), there are few
misclassifications, and the efficiency is close to unity. For variables for which the
discrimination is poor, such as the kinetic energy in figure \ref{fig:mynn}(c),
approximately half of the cells are misclassified, and the efficiency approaches the random
value 0.5.

The table in figure \ref{fig:mynn}(a) shows that the best discriminating variable is always
the enstrophy, and that modifying cells with strong vorticity has a stronger effect than
modifying weaker ones, confirming our initial guess that intense vortices are the dominant
structures in this flow.

\begin{figure}[t]
\centerline{%
\quad%
\includegraphics[height=.32\textwidth,clip]%
{\figpath 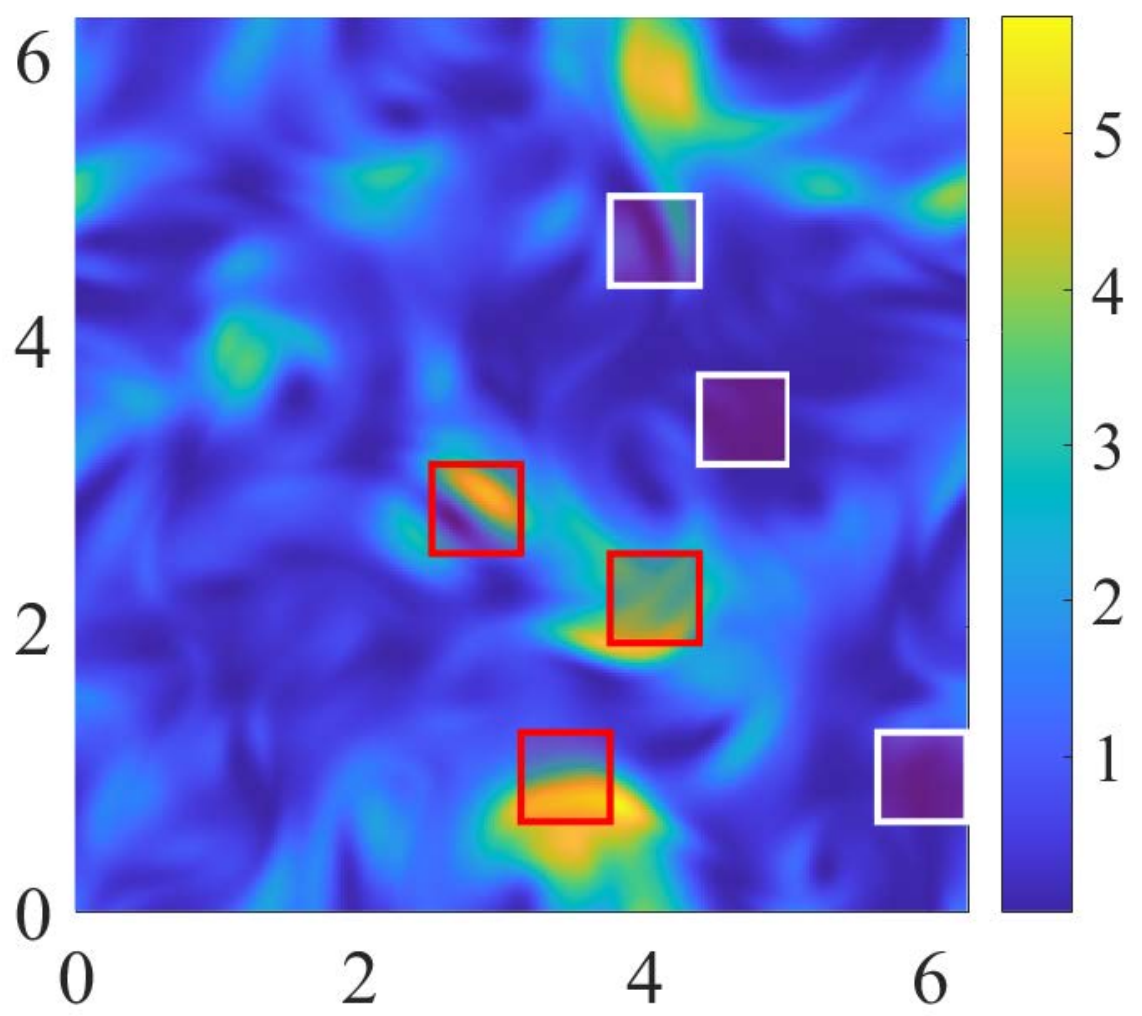}%
\mylab{-.40\textwidth}{.26\textwidth}{(a)}%
\hspace*{9mm}%
\raisebox{3.5mm}[0mm][0mm]{\includegraphics[height=.32\textwidth,clip]%
{\figpath 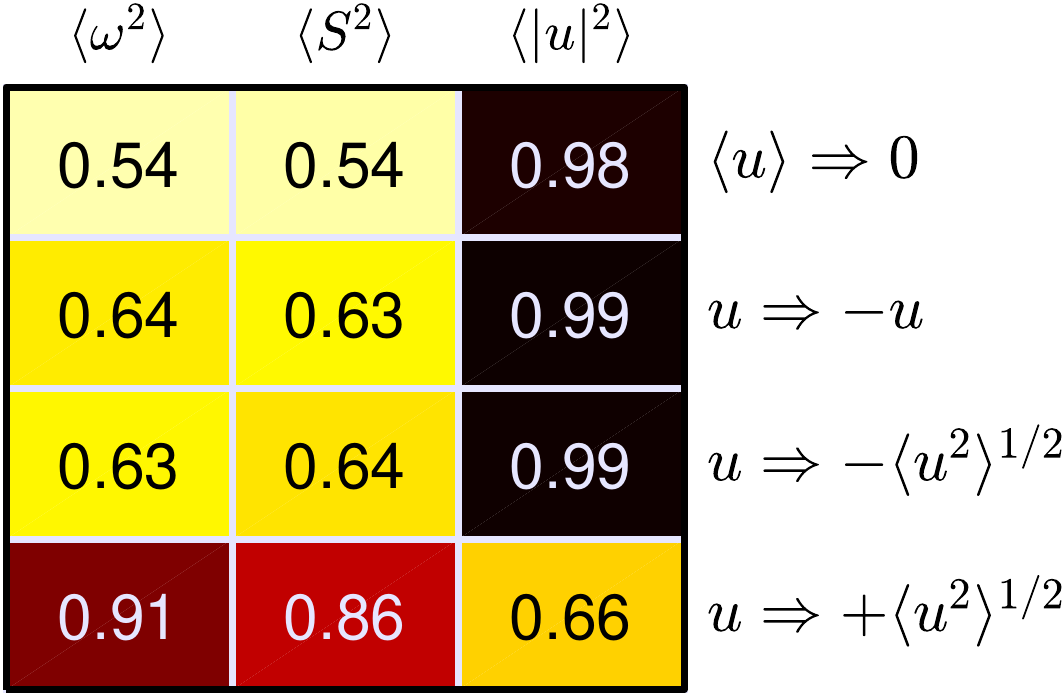}%
\mylab{-.54\textwidth}{.23\textwidth}{(b)}}%
}%
\caption{%
(a) Kinetic energy of an initial flow field to be tested, corresponding to the second line
from the bottom in (b). Outlined in white are the three least significant cells, and
in red are the three most significant ones. In this experiment the velocity in the
cell is substituted by a constant with opposite sign to its original mean, with a
magnitude that conserves the overall energy. Which velocity component is changed is chosen
to maximise the effect.
(b) Discrimination efficiency for a set of experiments that change the velocity of individual
cells of the flow in (a), as in figure \ref{fig:mynn}(a).
}
\label{fig:mynnu}
\end{figure}

However, figure \ref{fig:mynnu} suggests a more nuanced conclusion. Figure
\ref{fig:mynnu}(a) is a flow field from the same set of initial conditions as figure
\ref{fig:tur2d}, but the variable represented is the kinetic energy, which is also
intermittent in two-dimensional turbulence \cite{jimenez96}. The experiments in figure
\ref{fig:mynnu}(b) manipulate the velocity. This is more complicated than manipulating the
vorticity, because the modified velocity has to satisfy continuity (see \cite{jimploff18}
for details), and because velocity is a vector, and part of the experimental search is to
decide which component to modify. The table in figure \ref{fig:mynnu}(b) shows that the best
discriminating variable in these cases is the kinetic energy, not the enstrophy, suggesting
that there are at least two kinds of coherent structures in two-dimensional turbulence,
vortices and high-speed regions, and that they can be identified by different experimental
procedures.

Continuing with the interpretation of the above results is beyond the scope of the present
paper, whose subject is how computers can contribute to the study of turbulence, not the
dynamics of two-dimensional turbulence. But it is interesting that, even in this simple
case, random experiments unearth a set of structures that were a priori unexpected. Such
results can be considered examples of what we denoted above as machine-generated questions.
For example, it is known that high velocities are associated with vortex cores in
two-dimensional turbulence \cite{jimenez96}, suggesting that the two types of features may
not be that different, but this does not explain why the optimum discriminating variable is
different in the two cases. It is also interesting to highlight the bottom row in figure
\ref{fig:mynnu}(b), for which the best discriminating variables are the enstrophy and the rate of
strain, rather than the energy, even if it represents a velocity manipulation which only
differs from the next row by a sign.

In practice, the number of discriminating variables and experimental procedures may be quite
higher than in the previous example, and simple threshold discriminators may not be
sufficient for a successful classification. There are machine learning algorithms to handle
these cases, typically under the name of ensemble learning \cite{Rok:10}. Even in the
previous example, the joint use of enstrophy and energy as discriminating variables, which
can still be done by hand, results in a somewhat better discrimination than the single
variables used above. There are also automatic schemes for detecting outliers
\cite{Hodge:04}, such as the last line in figure \ref{fig:mynnu}(b). Their purpose is
often to eliminate them as noise, but, in the present context, it may rather be to
highlight interesting cases.

An important aspect of the experiments just described is cost, because trivial turbulence
simulations have only recently become possible. Our example uses a $256^2$ grid, and each
simulation spanning $\omega' T=6$ takes about 9\,s in a single Xeon core (X5650). Testing
each $10\times 10$ partition requires 100 simulations, and each experiment is run 100 times.
Each full experiment thus cost 37 core-hours. How many target times and experimental
manipulations are tested is up to the researcher. In the present case, it adds another
factor of 100, bringing the total to a few thousand core-hours. However, simulations are
independent of each other and can be run in parallel, and even a small departmental cluster
brings the actual `clock' time to about a week.

To translate these times to three-dimensional isotropic turbulence, consider a $256^3$ box.
Simulating $u'T/L=2$ on a modern GPU (Titan V) takes about two minutes, so that testing a
partition of $5^3$ cells requires approximately four GPU-hours. Assuming again 100
independent initial conditions brings the cost of a basic experimental test to 15 GPU-days.
Again, a small GPU cluster brings it down to a few days, and allows $O(10)$ randomised
experiments in a month. While more expensive than the two-dimensional case, this is still
affordable.

\section{Discussion: competitors or colleagues?}\la{sec:discus}

As mentioned in the introduction, the goal of this paper is to promote discussion, and it
therefore offers no conclusions. However, we can summarise what we have discussed up to now,
and highlight some of the questions that, from the point of view of the author, need to
be addressed by the community. The paper has been organised in three sections ordered by the
increasing capabilities of computers. The first two sections, which deal with computers as
expensive experimental apparatus, or as providers of routine answers to questions posed by
the researcher, present few conceptual problems. They represent different ways of doing
science as usual. As with all new instruments, computers have led to new discoveries that
would have been impossible had they not been available, and we have highlighted some of
them. In particular, the availability of four-dimensional data sets, which can be explored
in any desired order, and the possibility of performing almost arbitrary conceptual
experiments, have advanced our understanding of turbulence well beyond what was available in
the 1970's. The methodology discussed in these two sections can be considered as
established.

Section \ref{sec:curio} presents a different problem. In essence, the methods discussed
there have not yet achieved anything, but they promise to free turbulence research from
possible prejudices of the past, and to lead it to fresh territories. In the absence of an
ideal `very smart graduate student from Mars', who knows mathematics but does not know
enough terrestrial fluid mechanics to be misled by any errors that may have crept into our
theoretical models, the suggestion is that we may be able to outsource at least part of our
exploratory work to machines which, although not particularly smart, can deal with the
Navier--Stokes equations in trivially short times.

In the nomenclature popularised by Kuhn \cite{kuhn70}, machines do not share the paradigms
of trained scientists, and are freer to explore new avenues. While it would be unreasonable
to expect what we have called Monte Carlo experimentation to change the paradigms of our
discipline, one of the routine roles of research is to collect facts related to the
established point of view. The ostensible goal is to reinforce it, but history shows that,
occasionally, the resulting observations contradict the established theory, and that the
accumulation of such `mistakes' may result in an eventual paradigm shift, or at least in
paradigm drift. The hope is that the Monte Carlo research made possible by plentiful
computer power can be used as a tool to expedite this process, without prejudging the
result, in much the same way as we use machine tools to expedite manufacturing. In essence,
if we consider paradigms as optima of understanding, computers can be used to provide the
`noise' required to prevent us from getting stuck in a local optimum.
 
However, we cannot avoid worrying about whether the introduction of this new tool may be
different from the previous mechanical ones, and whether it may happen that we do not like
the results of machine research. In essence, we have to decide whether we want to cooperate
with the machines in this area, or to compete with them.

For example, what about the value of the K\'arm\'an constant? We mentioned in \S
\ref{sec:curio} that we can compute it quite precisely by blind simulation. Should we allow
the machine to conclude that knowing the value of $\kappa$ is the final answer? Should we
agree that a better turbulence theory is one that produces a more precise value for that
constant? If not, why not? One of the central tenets of science is that the results should
be independent of the feelings of the scientist, and it is doubtful from a `moral'
perspective whether we should reject useful science, such as a new antibiotic or
a better value for a modelling constant, just because we don't like the way in which it
was obtained.

But we also know, or at least this author does, that mathematics are beautiful, and that we
may be reluctant to accept, or to be interested in, ugly science. We may need to imbue
machines with a sense of aesthetics to make sure that we want to talk to them, or we risk
splitting science into useful and beautiful; not necessarily contradictory, but
neither necessarily coincident. The problem is more acute if we accept the argument in
\S\ref{sec:curio} that the results of randomised experiments are tantamount to questions
rather than to answers. Would we be as interested in machine questions as we are in machine
answers? Would machine curiosity coincide with human one? Would we even recognise it as
such?

Of course, most of these worries are overblown, at least in the short term, because machines
are still very far from matching even the most hapless laboratory assistant, but it may not
be too early to start discussing how to respond to them when they improve, as they will. Humanity
has been through this process before. Javelin throw used to be a useful military skill in
which champions were expected to throw as far as possible. It has since been relegated to a
sporting event, widely admired but full of restrictive rules intended to prevent `ugly' long
throws. Long-range attacks have been outsourced to machines.

We cited at the beginning of \S\ref{sec:structures} a well-known paper whose title asks
whether we would be able to understand what is it like to be a bat \cite{Nagel:74}. The answer in
that paper is not very hopeful. The question facing us in the medium term is whether we
would be able to recognise a machine as a `colleague'.

\vspace{2ex}
This work was supported by the European Research Council under the Coturb grant
ERC-2014.AdG-669505. 
%

\end{document}